\documentclass[a4paper,11pt]{article}
\usepackage{jheppub} % for details on the use of the package, please
\usepackage{bm}
\usepackage{soul}

\usepackage[T1]{fontenc} % if 

\newcommand\Schrodinger {Schr\"{o}dinger }

\newcommand\Ac {{\cal A}}

\DeclareMathOperator{\csch}{csch}

\title{\boldmath Boundary Conditions and Entanglement
	in Anti-de Sitter Space}

\author[a]{K. Boutivas,}
\author[a]{D. Katsinis,}
\author[a,b]{G. Pastras,}
\author[a,c]{N. Smeyers}
\author[a]{and N. Tetradis}

\affiliation[a]{Department of Physics, University of Athens, Zographou 157 84, Greece}
\affiliation[b]{Laboratory for Manufacturing Systems and Automation, Department of Mechanical Engineering and Aeronautics, University of Patras, Patra 26504, Greece}
\affiliation[c]{D\'epartement de Physique, \'Ecole Normale Sup\'erieure, 75005 Paris, France}

\emailAdd{kboutivas@phys.uoa.gr}
\emailAdd{dkatsinis@phys.uoa.gr}
\emailAdd{pastras@lms.mech.upatras.gr}
\emailAdd{nsmeyers@clipper.ens.psl.eu}
\emailAdd{ntetrad@phys.uoa.gr}

\abstract{We study the entanglement entropy of a 
	conformally coupled scalar field at its ground state in $(3+1)$-dimensional AdS space in global coordinates.
	 We consider spherical entangling surfaces centered at the origin of AdS and allow for general boundary conditions at the conformal boundary.
	 Through numerical and analytical means, we show that
	 the UV-divergent part of the entropy has a universal form, while
	 the UV-finite part is sensitive to the boundary conditions.
	 We determine the
	 dependence of the latter part on mixed boundary 
	 conditions that interpolate between Dirichlet and Neumann. }

\begin{document} 
\maketitle
\flushbottom

\section{Introduction} 
\label{introduction}

The entanglement entropy of quantum fields in $3+1$ dimensions has some remarkable properties that have attracted a lot of attention over the years. Its most famous feature, apparent in the simple case of a free scalar field in flat space at the ground state, is that it is dominated by an area-law term, similarly to the black hole entropy \cite{Sorkin:1984kjy,Bombelli:1986rw,Srednicki:1993im}. This leading contribution is proportional to $R^2/\epsilon^2$, where $R$ is the radius of the entangling surface and $\epsilon$ a cutoff regulating the ultraviolet (UV) divergence that results from the strong entanglement of neighbouring degrees of freedom across this surface. The identification of the UV cutoff with the energy scale at which the effective quantum field theory description breaks down is natural.
%, as new physics is expected to emerge beyond this scale. 
If one wishes to interpret the entanglement entropy as the microscopic origin of black-hole entropy, the natural choice for this cutoff is the Planck scale. However, establishing this correspondence quantitatively requires  treating gravity beyond the level of a fixed external background that does not respond to the presence of the quantum field. 

The first subleading term in the entropy is logarithmic, of the form $ \ln(R/\epsilon)$, with a universal coefficient equal to $-1/90$ for the scalar theory \cite{Solodukhin:2008dh,Casini:2010kt,Casini:2009sr,Lohmayer:2009sq}. This is related to the coefficient of the A-type conformal anomaly expressed in terms of the Euler density \cite{Solodukhin:2008dh}. The UV-finite part is of high interest as well. It has been used in order to define an $a$-function \cite{Cardy:1988cwa,Jack:1990eb,Komargodski:2011vj,Hartman:2023qdn}, which must interpolate between the coefficients of the $A$-type conformal anomaly for the theories at the two ends of the renormalization-group (RG) flow. The RG running is realized through the dependence of an appropriate function deduced from the entanglement entropy on the entangling radius $R$ \cite{Casini:2004bw,Casini:2005zv,Casini:2012ei,Casini:2017vbe}. Explicit calculations in simple models confirm the validity of this proposal \cite{Boutivas:2025qan,Abate:2026apg}.

The above picture can be extended to field theories on gravitational backgrounds. Typical cases include de Sitter (dS) space \cite{Maldacena:2012xp,Boutivas:2024sat,Boutivas:2024lts}, the Einstein universe \cite{Boutivas:2025rdf}, more general Friedemann-Robertson-Walker (FRW) cosmologies \cite{Boutivas:2023ksg,Boutivas:2023mfg}, and anti-de Sitter (AdS) space \cite{Boutivas:2025ksp}. The basic properties of the entropy persist to all the above cases. The leading contribution is given by a UV-divergent term proportional to the proper area of the entangling surface, while, for conformal theories, the coefficient of the logarithmic term is linked to the A-type conformal anomaly \cite{Solodukhin:2008dh}. The definition of appropriate $a$-functions can be extended to the dS \cite{Abate:2024nyh} and the AdS case \cite{Abate:2026apg}.

Several novel features have also been discovered: 
\begin{itemize}
\item In the presence of a gravitational background, an additional logarithmic UV divergence 
appears. Its coefficient is proportional to the area of the entangling surface in units of the characteristic length scale of the background \cite{Boutivas:2024sat,Boutivas:2024lts,Boutivas:2025rdf,Boutivas:2025ksp}. A similar divergence is present for massive fields, even in flat space, with the length scale set by the inverse mass \cite{Hertzberg:2010uv}. The two contributions cancel each other for a scalar field whose mass results from a conformal coupling to gravity. In this case, the only logarithmic divergence is linked to the conformal anomaly. 
\item Certain backgrounds enhance the sensitivity of the entanglement entropy to large length scales, an infrared (IR) phenomenon particularly amplified in the presence of zero modes. Corresponding IR-sensitive terms in the entropy have been identified in dS space \cite{Boutivas:2024sat,Boutivas:2024lts} and for
the Einstein universe \cite{Boutivas:2025rdf}.
\item The area law is a typical feature for fields at their ground state on highly symmetric backgrounds. In more general cases, such as during the transition from a quasi-dS phase to a radiation- or matter-dominated phase for a general FRW background, a volume term can appear, as is typically the case during a quench of the system \cite{Boutivas:2023ksg,Boutivas:2023mfg}.
\end{itemize}

The purpose of the present work is to investigate another aspect of the entanglement entropy of quantum fields on curved backgrounds, namely its possible dependence on the choice of boundary conditions. 
In AdS space, the presence of the conformal boundary allows for a family of admissible boundary conditions within the Breitenlohner-Freedman window \cite{Breitenlohner:1982bm,Breitenlohner:1982jf}, 
a range of negative values of the squared mass of the field, down to a specific value beyond which the theory is unstable.
It is possible that the choice of boundary conditions affects the entanglement entropy even for spherical regions near the center of AdS. 
We emphasize that we do not consider a boundary cutting off part of the space at some proper radial distance. We explore instead the dependence of the entanglement entropy on the asymptotic form of the field, for the parameter range in which this is not fixed uniquely.

For a spherical entangling surface of proper area $\Ac$, the entanglement entropy in AdS space is expected to include the terms
\begin{equation}
	S_{\textrm{AdS}}=\frac{d_1}{4\pi}\frac{\Ac}{\epsilon^2}+
	\left(d_2 + d_3 \frac{\Ac}{4\pi} \mu^2 + d_4 \frac{\Ac}{4\pi a^2} \right)\ln\frac{a}{ \epsilon}  
	+ d_5 \frac{\Ac}{4\pi a^2}  + \frac{d_6}{2} \ln\left(\frac{\Ac}{a}\right) +\cdots \label{eq:SEE_expansion_mald}
\end{equation}
where we have used the AdS length $a$ in order to define dimensionless quantities. Most of the constants in the above expression have known values for Dirichlet boundary conditions \cite{Boutivas:2025ksp}. The leading UV-divergence is proportional to a scheme-dependent constant $d_1$. For a regularization through the lattice discretization of the radial direction, introduced in the seminal work of Srednicki \cite{Srednicki:1993im}, we have $d_1\simeq 0.295$. The constants in the logarithmic UV-divergent term are $d_2=-1/90$, $d_3=-1/6$, $d_4=-1/3$. The constants in the finite part have not been computed before, even though consistency with the flat-space limit requires $d_6=-1/90$. The main objective of the present work is to determine how the choice of boundary conditions modifies these coefficients, beginning with the leading UV-divergent contributions and proceeding to the finite terms.
This study is a direct continuation of our previous work \cite{Boutivas:2025ksp}, in which the issue of boundary conditions was not addressed in full detail. In particular, we are interested in field masses for which several boundary conditions are available.

The interesting range of masses includes the conformal point, corresponding to the mass generated through a conformal coupling of the scalar field to gravity. At this point the entropy takes a particularly simple form, as the terms proportional to $d_3$ and $d_4$ in equation (\ref{eq:SEE_expansion_mald}) cancel. As a result, the effect of the boundary conditions is isolated from additional physical scales introduced by either the mass or the curvature. This case is also of special interest because the theory remains conformal for Dirichlet and Neumann boundary conditions, while mixed boundary conditions introduce a new scale, whose variation leads to an 
interpolation between the two conformal theories.

Finally, it is a fortunate coincidence that the numerical analysis is considerably simplified at the conformal point. As will become clear in the following section, imposing general mixed boundary conditions requires controlling the asymptotic expansion of the field near the conformal boundary. For the conformal mass this expansion becomes regular, making the numerical calculation of the different solutions much more straightforward.
  
In Section~\ref{ads}, we summarize the general setup for the analysis. We present the framework for imposing mixed boundary conditions in AdS space, the method of discretization used for the numerical study, and the formalism for computing the entropy. In Section~\ref{sec:numresults} we summarize the methodology and present the results, which are discussed in the final section. In the appendix we present the analytical calculation of the entanglement entropy of the $(1+1)$-dimensional massless theory with Dirichlet and Neumann boundary conditions at the two ends of a segment.
The analytical results confirm aspects of the numerical study 
presented in the main text.

\section{General Setup} 
\label{ads}
\subsection{Basic framework}  

The metric of AdS$_{d+1}$ space in global coordinates reads
\begin{equation}
	ds^2
	=
	-f(r)dt^2
	+\frac{1}{f(r)}dr^2
	+r^2d\Omega^2_{d-1},
	\quad 
	f(r)
	=
	1+\frac{r^2}{a^2},
	\label{eq:globalr}
\end{equation}
where $a$ is the AdS length. We study the entanglement entropy of a free scalar field at its ground state, when a spherical region around the origin of AdS is traced out at a given time. The conformal compactification of global AdS is obtained by introducing the tortoise coordinate $w$ according to 
\begin{equation}
	r
	=
	a\tan \frac{w}{a}\,,
	\label{eq:tortoise}
\end{equation}
so that $w=0$ corresponds to the center of AdS and $w=a\pi/2$ to the conformal boundary. In terms of the tortoise coordinate, the metric reads
\begin{equation}
	ds^2
	=
	\frac{1}{\cos^2\frac{w}{a}}
	\left(
	-dt^2
	+dw^2
	+a^2\sin^2\frac{w}{a} d\Omega^2_{d-1}
	\right).
	\label{eq:globalw}
\end{equation}
The action of the scalar field is
\begin{equation}
	\mathcal{S}
	=
	\frac{1}{2}\int dt \int_0^{a\frac{\pi}{2}} \hspace{-0.25cm} dw\int_{\textrm{S}^{d-1}} \hspace{-0.25cm} d\Omega_{d-1}
	\left(a\tan\frac{w}{a}\right)^{d-1}
	\left[
	\dot{\phi}^2
	-\left(\partial_w\phi\right)^2
	+\frac{\phi\Delta_{d-1}\phi}{a^2\sin^2\frac{w}{a}}
	-\frac{\mu^2\phi^2}{\cos^2\frac{w}{a}}
	\right],
	\label{eq:ScalarAction}
\end{equation}
where a dot denotes a derivative with respect to time, $\Delta_{d-1}$ is the Laplacian on the unit $(d-1)$-dimensional sphere, and $\mu$ the mass of the field. The field is expanded in terms of real hyper-spherical harmonics as
\begin{equation}
	\phi\left(t,w,\hat{r}\right)
	=
	\frac{1}{a^{\frac{d-1}{2}}\tan^{\frac{d-1}{2}}\frac{w}{a}}
	\sum_{\ell,\vec{m}}\phi_{\ell\vec{m}}\left(t,w\right)Y_{\ell \vec{m}}\left(\hat{r}\right),
	\label{eq:mode_expansion}
\end{equation}
where $\Delta_{d-1}Y_{\ell \vec{m}}\left(\hat{r}\right)=\ell(\ell+1)Y_{\ell \vec{m}}\left(\hat{r}\right)$, $\vec{m}=(m_1,\dots,m_{d-2})$ and $\hat{r}$ is a unit vector. In this way the original theory is divided into angular-momentum sectors that do not interact with each other,
\begin{equation}
	\mathcal{S}
	=
	\sum_{\ell,\vec{m}} \mathcal{S}_{\ell\vec{m}} ,
\end{equation}
where
\begin{multline}
	\mathcal{S}_{\ell\vec{m}}
	=
	\frac{1}{2}\int dt \int_0^{a\frac{\pi}{2}} dw 
	\left[
	\dot{\phi}_{\ell\vec{m}}^2
	-\left(\partial_w\phi_{\ell\vec{m}}\right)^2
	-\frac{1}{a^2}\left(\frac{\nu^2-\frac{1}{4}}{\sin^2\frac{w}{a}}+\frac{\kappa^2-\frac{1}{4}}{\cos^2\frac{w}{a}}\right)\phi^2_{\ell\vec{m}}
	\right]
	\\
	+\frac{d-1}{2}\int dt \int_0^{a\frac{\pi}{2}} dw\, 
	\partial_w\left(\frac{\phi^2_{\ell\vec{m}}}{a\sin\frac{2w}{a}}\right),
	\label{eq:Action_lm}
\end{multline}
with the parameters $\nu$ and $\kappa$ given by
\begin{equation}
	\nu
	=
	\ell+\frac{d}{2}-1\,,
	\quad 
	\kappa
	=
	\sqrt{\mu^2 a^2+\frac{d^2}{4}}\,.
	\label{eq:nu_kappa_def}
\end{equation}

The definition of the physical modes requires specific conditions for the behavior of $\phi_{\ell\vec{m}}$ at the points $0$ and $a\pi/2$. These conditions imply that the original field $\phi$ is regular at the origin and obeys particular boundary conditions at the conformal boundary. How these translate to specific conditions for the moments $\phi_{\ell\vec{m}}$ is described in detail in \cite{Boutivas:2025ksp}. Instead of repeating this discussion here, we summarize only the points that are relevant for our calculation. 

Once suitable boundary conditions have been imposed, the variational problem leads to the following second order equation for the modes $\phi_{\ell\vec{m}}$  
\begin{equation}
	\ddot{\phi}_{\ell\vec{m}}
	-\partial^2_w\phi_{\ell\vec{m}}
	+\frac{1}{a^2}\left(\frac{\nu^2-\frac{1}{4}}{\sin^2\!\frac{w}{a}}+\frac{\kappa^2-\frac{1}{4}}{\cos^2\!\frac{w}{a}}\right)\phi_{\ell\vec{m}}
	=
	0\,.
\end{equation}
%with
%\begin{equation}
%	\nu
%	=
%	\ell+\frac{d}{2}-1\,,
%	\qquad 
%	\kappa
%	=
%	\sqrt{\mu^2 a^2+\frac{d^2}{4}}\,.
%	\label{eq:nu_kappa_def}
%\end{equation}
Separation of variables as $\phi_{\ell\vec{m}}(t,w)=e^{\pm iEt} \varphi_{\ell\vec{m}}(w)$ gives
\begin{equation}
	-\partial^2_w\varphi_{\ell\vec{m}}
	+\frac{1}{a^2}\left(\frac{\nu^2-\frac{1}{4}}{\sin^2\!\frac{w}{a}}+\frac{\kappa^2-\frac{1}{4}}{\cos^2\!\frac{w}{a}}\right)\varphi_{\ell\vec{m}} 
	= 
	E^2 \varphi_{\ell\vec{m}}\,.
	\label{eq:eigensystem}
\end{equation}
The unique solution of this equation that satisfies the regularity condition at $w=0$ is\footnote{We have shifted the definition of $s$ by 1 compared to \cite{Boutivas:2025ksp}.}
\begin{equation}
	\varphi_{\ell\vec{m}}(w;s)
	=C\,
	%\frac{c_s}{\sqrt{a}}\frac{\Gamma(\nu+s+1)}{\Gamma(s+1)\Gamma(\nu+1)}
	 \sin^{\nu+\frac{1}{2}}\!\frac{w}{a}
	 \cos^{\kappa+\frac{1}{2}}\!\frac{w}{a}\,
	 {}_2F_1\left(-s+1,s+\kappa+\nu,\nu+1,\sin^2\!\frac{w}{a}\right),
	 \label{eq:spectum_K}
\end{equation}
where ${}_2F_1(\alpha,\beta,\gamma,z)$ is the Gauss hypergeometric function and $s$ a spectral parameter that determines the energy via the relation through
\begin{equation}
	E_s 
	= 
	\frac{1}{a} \left(2s+\kappa+\nu-1\right)\,.
	\label{eq:eigenvalues}
\end{equation}
The normalization constant $C$ depends on the parameters of the solution, but its exact value is not important for the following.

The behavior of the above solution near the boundary reflects the
possible boundary conditions. Expanding the solution for $w^\prime= a\frac{\pi}{2}-w\ll a$ gives
\begin{multline}
	\varphi_{\ell\vec{m}}(w;s) 
	\sim 
	%=-\frac{c_s}{\sqrt{a}}\frac{1}{\sin(\pi\kappa)}\left[
	\frac{\pi\left(\frac{w^\prime}{a} \right)^{\kappa+\frac{1}{2}}}{\Gamma(s)\Gamma(-s-\kappa+1)\,\Gamma(\kappa+1)}+{\mathcal{ O}}\left(\left(\frac{w^\prime}{a}\right)^{\kappa+\frac{5}{2}}\right)
	% \right. 
	\\
	%\left.
	-\frac{\sin(s \pi)\,\Gamma(s+\nu)\left(\frac{w^\prime}{a} \right)^{-\kappa+\frac{1}{2}}}{\Gamma(1-\kappa)\,\Gamma(s+\kappa+\nu)}+{\mathcal{ O}}\left( \left(\frac{w^\prime}{a}\right)^{-\kappa+\frac{5}{2}}\right)
	%\right]
	\,.
	\label{eq:asympt1a} 
\end{multline}
This asymptotic form implies that the solution~\eqref{eq:spectum_K} is normalizable for any $\kappa>0$ provided $s$ is a positive integer\footnote{Setting $s \rightarrow 1-s$ does not generate new linearly independent solutions, but rather corresponds to the symmetry $\kappa \rightarrow -\kappa$, $\nu \rightarrow -\nu$.}, so that the second term vanishes. On the other hand, this term can be present in a normalizable solution for small positive values of $\kappa$. It turns out that the solution~\eqref{eq:spectum_K} is normalizable for any $s$ provided $0\leq\kappa< 1$ \cite{Ishibashi:2004wx}. This reflects the fact that the most general asymptotic behavior of the original scalar field $\phi$ for $\kappa >0$ takes the form
\begin{equation}
	\phi\left(t,w,\hat{r}\right)
	\sim 
	\phi_-(t,\hat{r})\left(a\frac{\pi}{2}-w\right)^{d-\Delta}
	+\cdots
	+\phi_+(t,\hat{r})\left(a\frac{\pi}{2}-w\right)^{\Delta}
	+\cdots\,,
	\label{eq:AdSasymptotics}
\end{equation}
where $\phi_\pm(t,\hat{r})$ are arbitrary functions and $\Delta=\kappa+d/2$. A special expression holds for $\kappa=0$ for which $d-\Delta = \Delta$, see for example \cite{Boutivas:2025ksp}.

The presence of two types of asymptotic behavior allows the implementation of a variety of boundary conditions, depending on the relative weight of the two terms in equation~\eqref{eq:asympt1a}. As analyzed in detail in \cite{Ishibashi:2004wx,Barroso:2019cwp,SAEbook}, the mixed boundary conditions that are admissible for $\kappa$ in the range $0\leq\kappa< 1$ correspond to a one-parameter family of self-adjoint extensions of the Hamiltonian appearing in equation~\eqref{eq:eigensystem} when viewed as an effective \Schrodinger equation.

On the other hand, the form of the expansion~\eqref{eq:asympt1a} demonstrates the difficulty of implementing these boundary conditions in a numerical study of the system. One has to fix not only the value of the field or its derivative at some point, but the full asymptotic form of the solution. A particularly convenient case for the numerical implementation is $\kappa=1/2$. As can be seen from~\eqref{eq:asympt1a}, for this value of $\kappa$ the field and its derivative have regular expansions around the conformal boundary. 

A field with a non-minimal coupling $\frac{1}{2}\xi R \phi^2$ in AdS develops an effective mass
\begin{equation}
	\mu^2_{\rm eff}
	=
	\mu^2-\xi\frac{d(d+1)}{a^2}\,.
\end{equation}
For $\xi=\frac{d-1}{4d}$, which corresponds to a Weyl-invariant theory, we have $\kappa=\sqrt{\mu^2a^2+\frac{1}{4}}$  for any value of $d$. Thus, a conformally coupled, massless scalar corresponds to $\kappa=1/2$. We focus on this case in the following, as it eliminates additional scales, such as the bare mass, from the problem and isolates the effect of the boundary conditions. It is also particularly convenient numerically, as we discussed above.

\begin{figure}[t]
	\centering
	\vspace{-1cm}
	\begin{picture}(92,30)
		\put(0,0){\includegraphics[width=0.4\textwidth]{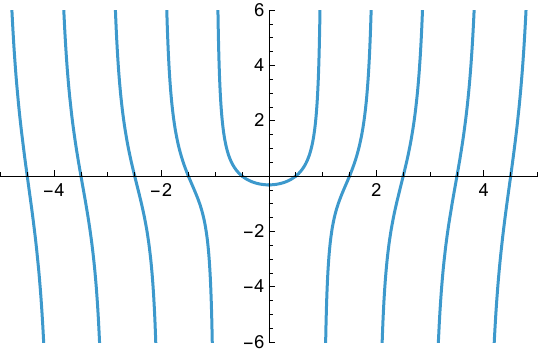} }
		\put(50,0.){\includegraphics[width=0.4\textwidth]{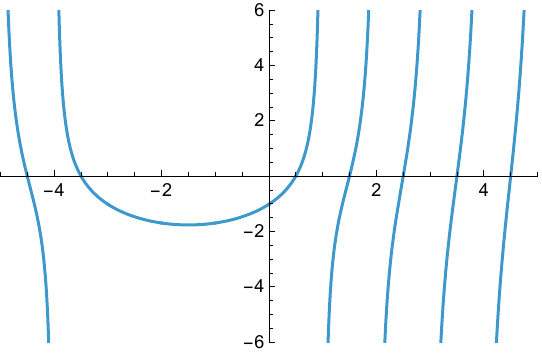} }
		\put(40,12.75){{$s$}}
		\put(90,12.75){{$s$}}
		\put(18.75,26.5){{$A$}}
		\put(68.75,26.5){{$A$}}
	\end{picture}
	\caption{Left plot: The function $A(s)$ defined in~\eqref{eq:gbc} for $\ell=0$. Right plot: The same function for $\ell=3$.}
	\label{fig:plot0}
\end{figure}
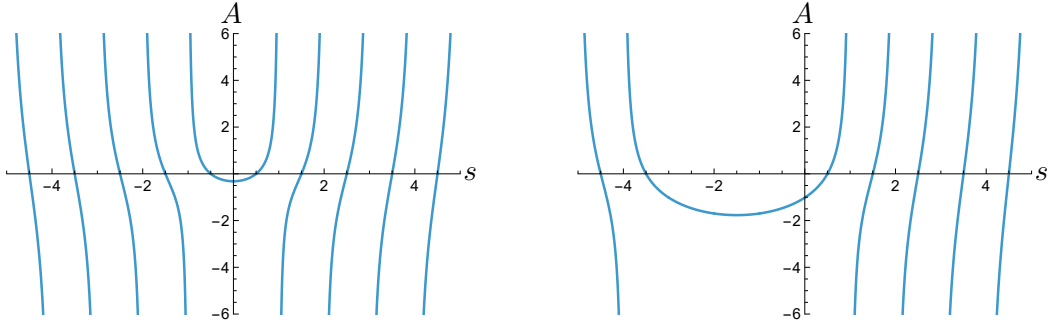

According to~\eqref{eq:asympt1a} with $\kappa=1/2$,
%and following the conventions of \cite{Boutivas:2025ksp},
a general boundary condition can be implemented by fixing the ratio
\begin{equation}
	A
	:=
	-\frac{a}{2} \left. \frac{\partial_w \varphi_{lm}(w,s)}{\phi_{lm}(w,s)}\right|_{w=a\frac{\pi}{2}}=
-\frac{\pi\, \Gamma\left[s+\nu+\frac{1}{2}\right]}{\sin(s \pi)\,\Gamma\left[\frac{1}{2}-s\right]\,
\Gamma[s]\,\Gamma[s+\nu]}.
	\label{eq:gbc}
\end{equation}
This is a specific implementation of the general framework of imposing boundary conditions in AdS space, discussed in detail in \cite{Boutivas:2025ksp}. The ratio $A$ as a function of $s$ is depicted in Fig.~\ref{fig:plot0} for two values of $\ell$. We are interested in positive values of $A$, which result in self-adjoint extensions of the effective \Schrodinger Hamiltonian that are positive-definite. Specific values of $A$ generate discrete spectra for the field modes. These can be read off from Fig.~\ref{fig:plot0} by drawing horizontal lines for fixed $A$. 
%For each angular-momentum $\ell$, a complete spectrum corresponds to the positive values of the spectral parameter $s$ at the points at which a line of given $A$ cuts the curves in Fig.~\ref{fig:plot0}. 
In the limit $A\to \infty$ the parameter $s$ takes positive integer values that lead to the vanishing of the sine-function in the denominator of~\eqref{eq:gbc}. This choice realizes Dirichlet boundary conditions. Requiring $A=0$ selects half-integer values of $s$ that correspond to the poles of $\Gamma\left[\frac{1}{2}-s\right]$ on the real axis. This choice realizes Neumann boundary conditions. Intermediate values of $A$ correspond to mixed boundary conditions. These do not lead to a regular pattern of eigenvalues, which makes an analytical treatment unfeasible.

For the remainder of this work, we restrict our attention to the $(3+1)$-dimensional spacetime case, fixing $d=3$. In this case, the hypergeometric function simplifies considerably; it reduces to Jacobi polynomials for $A\to \infty$ and Gegenbauer polynomials for $A=0$. This permits the explicit determination of the normalization factors and the direct comparison with the modes deduced from the numerical solutions in the discretized setup described in the following subsection. We have made numerous checks for many values of $A$, $\nu$ and $s$ in order to verify that the numerical solutions reproduce correctly the known analytical form of the field modes. 

\subsection{Discretization} 
\label{subsec:Discretization}

The calculation of the entropy is performed through the discretization of the system in the spirit of the seminal work of Srednicki \cite{Srednicki:1993im}. For the detailed description of the implementation in the case of an AdS background, we refer the reader to \cite{Boutivas:2025ksp}. The approach relies on descretizing the degrees of freedom of the field, resulting in a system of coupled harmonic oscillators that correspond to the field values on a radial lattice for each value of the angular momentum. When the lattice spacing is sent to zero, the eigenmodes of the discrete system must reproduce the modes found analytically in the previous section for given boundary conditions. This requires the introduction of appropriate boundary terms in the field Hamiltonian, which would select the correct eigenmodes. The numerical implementation, however, does not require the explicit specification of the boundary terms, as long as the form of the coupling matrix of the system of oscillators results in eigenmodes of the discretized system that reduce to the desired ones in the continuum limit. For this reason we will not discuss the structure of the boundary terms of the continuous theory explicitly. Instead, we will use a partial integration in order to write the bulk part of the Hamiltonian for $d=3$ and $\kappa=1/2$ as
\begin{equation}
	H_{\ell{m}}
	=
	\frac{1}{2}\int_0^{a\frac{\pi}{2}} dw 
	\left[
	\pi^2_{\ell{m}}
	-\phi_{\ell{m}}\partial^2_w\phi_{\ell{m}}
	+\frac{1}{a^2}\frac{\left(\ell+\frac{1}{2} \right)^2-\frac{1}{4}}{\sin^2\frac{w}{a}}\phi^2_{\ell{m}}
	\right]
	\label{eq:Hlm}
\end{equation}
for each angular-momentum sector, where $\pi_{\ell{m}}=\dot{\phi}_{\ell{m}}$. 
%and the boundary terms are left unspecified. 
For $ w\ll a$, the Hamiltonian assumes the same form as for a massless field in flat space. Thus the leading contributions to the entropy for entangling radii much smaller than the AdS length is expected to coincide with those on a flat background. 

The discretization of the theory is performed through the scheme 
\begin{equation}
	w
	=
	\epsilon\, i,
	\qquad 
	a
	=
	\frac{2}{\pi}\epsilon(N+1), 
	\qquad 
	\int_0^{a\frac{\pi}{2}} dw 
	\rightarrow 
	\epsilon\sum_{i=0}^{N+1},
	\label{eq:discretization}
\end{equation}
with the field modes redefined as
\begin{equation}
	\phi_{\ell{m}}(t,w)
	\rightarrow 
	\frac{1}{\sqrt{\epsilon}}\phi_{\ell{m},i}(t),
	\qquad 
	\pi_{\ell{m}}(t,w)
	\rightarrow 
	\frac{1}{\sqrt{\epsilon}}\pi_{\ell{m},i}(t).
\end{equation}
The dynamical variables are the oscillators $\phi_{\ell{m},i}$ with $i=1,\dots,N$, while the points $i=0$ and $i=N+1$ act as auxiliary in order to fix the field derivatives at the ends of the interval through differences. Expressing the second derivative with respect to $w$ at point $i$ as $(\phi_{\ell{m},i+1}+\phi_{\ell{m},i-1}-2\phi_{\ell{m},i})/\epsilon^2$ results in the expected tridiagonal form of the coupling matrix. However, care must be taken for the endpoints.

The regularity condition at $w=0$ is implemented  in the standard fashion through the assumption that the field vanishes at this point, so that $\phi_{\ell{m},0}=0$. On the other hand, the field value at the auxiliary point $i=N+1$ depends on the choice of boundary conditions. For Dirichlet we must assume $\phi_{\ell{m},N+1}=0$, while for Neumann $\phi_{\ell{m},N+1}=\phi_{\ell{m},N}$. For mixed boundary conditions, the first equality of~\eqref{eq:gbc} must be imposed for given $A$. Since the finite-difference approximation of the radial derivative is naturally associated with the last lattice link, an implementation of the mixed boundary condition is obtained by evaluating the field at the middle of the link, rather than at the last lattice site. This gives
\begin{equation}
	A
	=
	-\frac{a}{\epsilon} 
	\frac{\phi_{\ell{m},N+1}-\phi_{\ell{m},N}}{\phi_{\ell{m},N+1}+\phi_{\ell{m},N}},
	\label{eq:condd1} 
\end{equation}
which can be used in order to express $\phi_{\ell{m},N+1}$ in terms of $\phi_{\ell{m},N}$ and $A$. Variations of this condition are possible, but they only introduce modifications at
order $\epsilon^2$, with 
\begin{equation}
	\epsilon
	=
	\frac{\pi a}{2(N+1)}. 
	\label{eq:cutof} 
\end{equation}
We obtain $\phi_{\ell{m},N+1}=Q\, \phi_{\ell{m},N}$, with
\begin{equation}
	Q
	=
	\frac{1-\frac{\pi A}{2(N+1)}}{1+\frac{\pi A}{2(N+1)}}.
	\label{eq:Qdef}
\end{equation}
This discretization reproduces the Neumann boundary condition exactly for $A=0$, independently of the lattice size, as expected in the continuum theory. It also yields a Dirichlet boundary condition for a specific value $A=2(N+1)/\pi$ that depends on the
number of lattice points. This critical value approaches $A\to \infty$ for $N\to \infty$, thus
recovering correctly the continuum limit.

%The above relation has the expected form for Neumann boundary conditions, corresponding to $A=0$. The situation is not as clear for Dirichlet boundary conditions, corresponding to $A\to \infty$. This limit must be taken with care, making sure that the continuum limit is approached first with $N \gg A$. 

In this way, we end up with a system of $N$ interacting harmonic oscillators, with Hamiltonian
\begin{equation}
	H_{\ell{m}}
	=
	\frac{1}{2}\sum_{i=1}^N \pi^2_{\ell{m},i}
	+\frac{1}{2}\sum_{i,j=1}^N \phi_{\ell{m},i}\, K_{ij} \, \phi_{\ell{m},j} \, 
\end{equation}
and interactions determined by the coupling matrix
\begin{equation}
	K_{ij}
	=
	\frac{1}{a^2}
	\left[
	\frac{4(N+1)^2}{\pi^2}\left(2\delta_{i,j}-\delta_{i+1,j}-\delta_{i,j+1}-Q\,\delta_{i,N}\delta_{j,N}\right)
	+\frac{\left(\ell+\frac{1}{2} \right)^2-\frac{1}{4}}{\sin^2\frac{\pi i}{2(N+1)}}\delta_{i,j}
	\right].
	\label{eq:coupling matrix}
\end{equation}
In this formulation the AdS length $a$ can be used as the unit for dimensionful quantities. This allows us to set $a=1$ in the numerical calculation.

%\begin{figure}[t]
%	\centering
%	\vspace{-1cm}
%	\begin{picture}(90,30)
%		\put(0,0){\includegraphics[width=0.4\textwidth]{plot11l.pdf} }
%		\hspace{2cm}
%		\put(50,0.){\includegraphics[width=0.4\textwidth]{plot11r.pdf} }
%		\put(42,11.4){{\large $\frac{w}{a}$}}
%		\put(91.2,11.4){{\large $\frac{w}{a}$}}
%		\put(9.7,10){{\footnotesize $\frac{\pi}{8}$}}
%		\put(19.3,10){{\footnotesize $\frac{\pi}{4}$}}
%		\put(28.5,10){{\footnotesize $\frac{3\pi}{8}$}}
%		\put(38.5,10){{\footnotesize $\frac{\pi}{2}$}}
%		\put(59.6,10){{\footnotesize $\frac{\pi}{8}$}}
%		\put(69.3,10){{\footnotesize $\frac{\pi}{4}$}}
%		\put(78.4,10){{\footnotesize $\frac{3\pi}{8}$}}
%		\put(88.5,10){{\footnotesize $\frac{\pi}{2}$}}
%		\put(-1.1,4.){{\footnotesize $-1$}}
%		\put(0.5,19.3){{\footnotesize $1$}}
%		\put(49,4){{\footnotesize $-1$}}
%		\put(50.5,19.3){{\footnotesize $1$}}
%		%	\put(50.5,16.9){{\footnotesize $0.05$}}
%		%	\put(47.8,22.5){{\footnotesize $0.1$}}
%	\end{picture}
%	\caption{Left plot: The eigenvector of the coupling matrix (\ref{eq:coupling matrix}) with $N=300$, $A=50$, $\ell=5$	and the corresponding analytical mode of the continuous theory~\eqref{eq:spectum_K} (continuous line) with eigenvalue $s=3.97$. Right plot: The eigenvector of the coupling matrix~\eqref{eq:coupling matrix} with $N=300$, $Q=0$, $\ell=5$ and the corresponding analytical mode of the continuous theory~\eqref{eq:spectum_K} (continuous line) with eigenvalue $s=4$. We have set $a=1$.}
%	\label{fig:plot11}
%\end{figure}

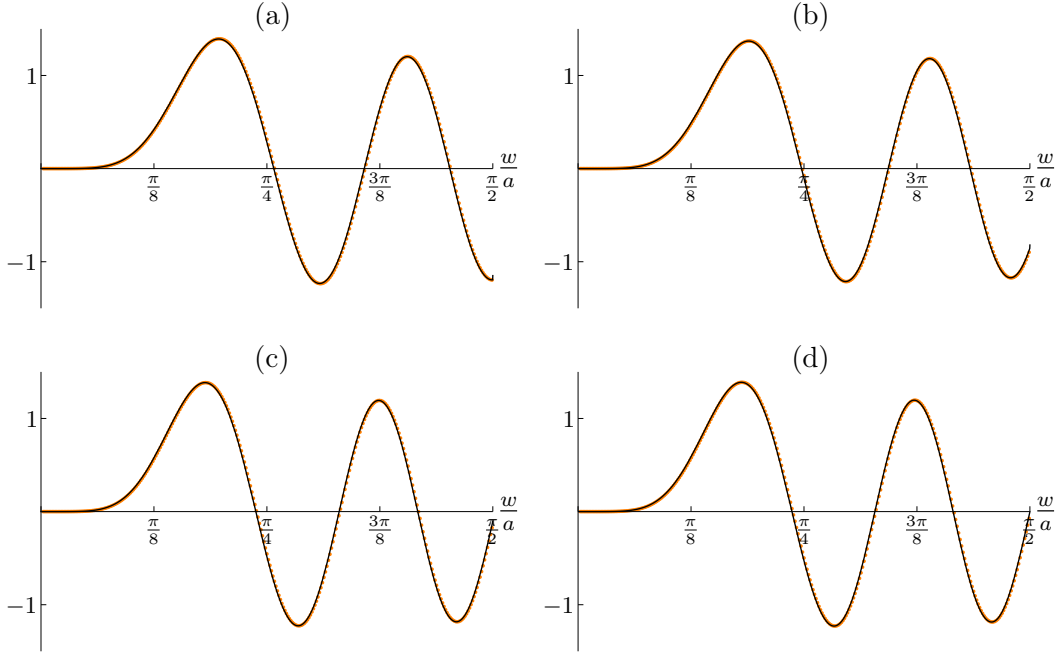
\begin{figure}[t]
	\centering
	\vspace{-1cm}
	\begin{picture}(94,58)
		\put(3,30){\includegraphics[width=0.4\textwidth]{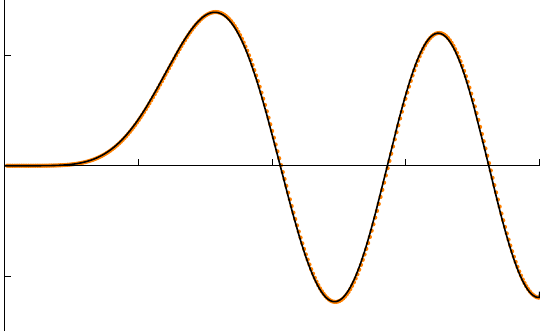} }
		\put(50,30.){\includegraphics[width=0.4\textwidth]{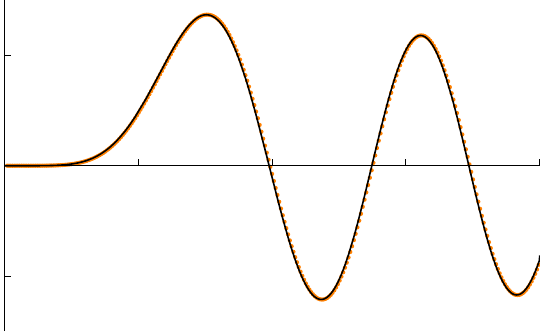} }
		\put(22,55){(a)}
		\put(69,55){(b)}
		\put(43.25,41.75){{\large $\frac{w}{a}$}}
		\put(12.375,40.25){{\footnotesize $\frac{\pi}{8}$}}
		\put(22.3,40.25){{\footnotesize $\frac{\pi}{4}$}}
		\put(31.75,40.25){{\footnotesize $\frac{3\pi}{8}$}}
		\put(42,40.25){{\footnotesize $\frac{\pi}{2}$}}
		\put(0.25,33.5){{\footnotesize $-1$}}
		\put(2,49.8){{\footnotesize $1$}}
		\put(90.24,41.75){{\large $\frac{w}{a}$}}
		\put(59.375,40.25){{\footnotesize $\frac{\pi}{8}$}}
		\put(69.3,40.25){{\footnotesize $\frac{\pi}{4}$}}
		\put(78.75,40.25){{\footnotesize $\frac{3\pi}{8}$}}
		\put(89,40.25){{\footnotesize $\frac{\pi}{2}$}}
		\put(47.25,33.5){{\footnotesize $-1$}}
		\put(49,49.8){{\footnotesize $1$}}
	%	\put(50.5,16.9){{\footnotesize $0.05$}}
	%	\put(47.8,22.5){{\footnotesize $0.1$}}
		\put(3,0){\includegraphics[width=0.4\textwidth]{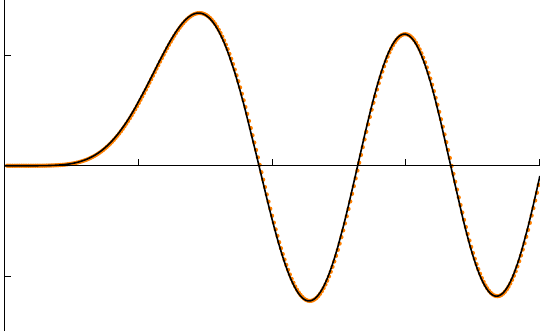} }
		\put(50,0){\includegraphics[width=0.4\textwidth]{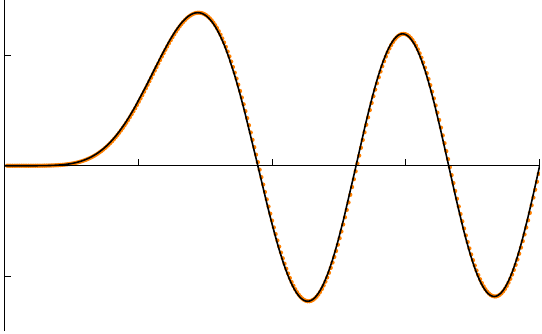} }
		\put(22,25){(c)}
		\put(69,25){(d)}
		\put(43.25,11.75){{\large $\frac{w}{a}$}}
		\put(12.375,10.25){{\footnotesize $\frac{\pi}{8}$}}
		\put(22.3,10.25){{\footnotesize $\frac{\pi}{4}$}}
		\put(31.75,10.25){{\footnotesize $\frac{3\pi}{8}$}}
		\put(42,10.25){{\footnotesize $\frac{\pi}{2}$}}
		\put(0.25,3.5){{\footnotesize $-1$}}
		\put(2,19.8){{\footnotesize $1$}}
		\put(90.24,11.75){{\large $\frac{w}{a}$}}
		\put(59.375,10.25){{\footnotesize $\frac{\pi}{8}$}}
		\put(69.3,10.25){{\footnotesize $\frac{\pi}{4}$}}
		\put(78.75,10.25){{\footnotesize $\frac{3\pi}{8}$}}
		\put(89,10.25){{\footnotesize $\frac{\pi}{2}$}}
		\put(47.25,3.5){{\footnotesize $-1$}}
		\put(49,19.8){{\footnotesize $1$}}
		%	\put(50.5,16.9){{\footnotesize $0.05$}}
		%	\put(47.8,22.5){{\footnotesize $0.1$}}
	\end{picture}
	\caption{The eigenvector of the coupling matrix~\eqref{eq:coupling matrix} with $N=300$, $\ell=5$ (orange dots) and the corresponding analytical mode of the continuous theory~\eqref{eq:spectum_K} (black curve).
	(a) The eigenvector with eigenvalue $s=3.5$ corresponding to $A=0$ (Neumann condition). (b) The eigenvector with eigenvalue $s=3.74$ corresponding to $A=5$. (c) The eigenvector with eigenvalue $s=3.97$ corresponding to $A=50$. (d) The eigenvector with eigenvalue $s=4$ corresponding to $A=2(N+1) / \pi$ in the discrete theory and $A = \infty$ in the continuous (Dirichlet condition). We have set $a=1$.}
	\label{fig:plot1}
\end{figure}

We have confirmed that the eigenmodes of the discrete Hamiltonian converge to the continuous eigenfunctions~\eqref{eq:spectum_K}. In Fig.~\ref{fig:plot1}, we display the eigenvectors of the coupling matrix with $N=300$ and the corresponding analytical modes for $\ell=5$ and different boundary conditions. Panel (a) depicts an eigenvector with eigenvalue $s=3.5$, satisfying a Neumann boundary condition with $A=0$. Panel (b) depicts an eigenvector with eigenvalue $s=3.74$, satisfying a mixed boundary condition with $A=5$ at $w=a\pi/2$. Panel (c) depicts an eigenvector with eigenvalue $s=3.97$, satisfying a mixed boundary condition with $A=50$ at $w=a\pi/2$. Finally, panel (d) depicts an eigenvector obtained by setting $Q=0$ directly in the coupling matrix~\eqref{eq:coupling matrix}. The agreement between the eigenvector of the discretized system (orange dots) and the analytical eigenfunction of the continuous system is apparent.

%As a final comment we would like to point out that, even though the eigenfunctions of the continuous system are known, they cannot be used directly in the context of the numerical calculation. A consistent approach based on the discretization of the system must employ a set of orthonormal eigenmodes equal in number to the degrees of freedom of the finite system. These can be computed numerically. Making use of a limited number of the infinite set of eigenmodes of the continuous system destroys the consistency, as they do not even form a complete basis. As a result, the latter eigenmodes can be used only in order to confirm the validity of the eigenmodes of the discrete system for low eigenvalues in units of the lattice spacing that acts as a UV cutoff.

\subsection{Calculation of the entropy}

The calculation of the entanglement entropy proceeds similarly to the analysis of Srednicki for a scalar field in flat space \cite{Srednicki:1993im}. The ground state wave function of the discretized system for each $(\ell,{m})$-sector is
\begin{equation}
	\Psi({{\bm \phi}_{\ell{m}}})
	=
	\left(\det\frac{\Omega}{\pi}\right)^{1/4} 
	e^{-\frac{1}{2} {\bm \phi}_{\ell{m}}^T \Omega\, {\bm \phi}_{\ell{m}}},
\end{equation}
where $\Omega$ is the positive square root of the coupling matrix $K$ given by~\eqref{eq:coupling matrix}, and ${\bm \phi}_{\ell{m}}$ is the column vector composed of the $\phi_{\ell{m},j}$. We consider the oscillators $1$ to $n$ as subsystem $A$, and $n+1$ to $N$ as the complementary subsystem $C$. The entanglement entropy is
\begin{equation}
	S_{\ell{m}}^{\textrm{EE}}
	=
	\sum_{i=1}^{N-n}
	\left(
	\frac{\sqrt{\lambda_i}+1}{2}\ln\frac{\sqrt{\lambda_i}+1}{2}
	-\frac{\sqrt{\lambda_i}-1}{2}\ln\frac{\sqrt{\lambda_i}-1}{2}
	\right),
	\label{eq:entropylambda}
\end{equation}
where $\lambda_i$ are the eigenvalues of the matrix
\begin{equation}
	\mathcal{M}
	=
	\left(\Omega^{-1}\right)_C\left(\Omega\right)_C.
	\label{eq:mmatrix}
\end{equation}
The matrix $\left(\Omega\right)_C$ is the $(N-n)\times(N-n)$ bottom-right block of $\Omega$ and similarly $\left(\Omega^{-1}\right)_C$ is the $(N-n)\times(N-n)$ bottom-right block of $\Omega^{-1}$. This result can be obtained by implementing the method of correlation functions for the calculation of the entanglement entropy \cite{Peschel:2002yqj} (see also \cite{Katsinis:2024gef}), which is equivalent to the approach of \cite{Srednicki:1993im}. 
%(See \cite{Katsinis:2024gef,Katsinis:2023hqn} for additional information on this equivalence.) 
The overall entanglement entropy is obtained by summing the contributions of all the $(\ell,{m})$-sectors, taking into account the degeneracy of these contributions. For AdS$_4$, i.e., for $d=3$, we have
\begin{equation}
	S_{\textrm{EE}}
	=
	\sum_{\ell=0}^\infty(2\ell+1)S_{\ell}^{\textrm{EE}}.
	\label{eq:SEE_total}
\end{equation}
Due to the spherical symmetry, $S_{\ell{m}}^{\textrm{EE}}$ does not depend on $m$; $S_{\ell}^{\textrm{EE}}$ denotes the contribution of a single $(\ell,{m})$-sector for any value of $m$.

%For $d>3$ the sum has to be regularized, as it is divergent.

\section{Numerical Analysis and Results }
\label{sec:numresults}

\subsection{Methodology}

According to the discretization scheme~\eqref{eq:discretization}, the degrees of freedom are the field values at  the positions $w = w_i$, with
\begin{equation}
	\frac{w_i}{a}=\frac{\pi}{2}\frac{i}{N+1},\qquad i=1,\dots , N .
\end{equation} 
We consider various radial lattices with $N = 49 + 50k$ for $k=0,\dots,9$. For each value of $N$, we compute the entanglement entropy for entangling surfaces with radii $w_n$,
where $n=(k+1)j$  with $j=1,\dots , 48 $,
%given by
%\begin{equation}
%	\frac{w_j}{a}=\frac{\pi}{2}\frac{j}{50},\qquad j=1,\dots , 48 ,
%\end{equation}
so that the entangling surfaces have the same physical radius for all $N$. This is required for the direct comparison of the entropies for various values of the UV cutoff. 

The calculation is repeated for several values of the parameter $A$ that determines the boundary conditions on the conformal boundary for the various field modes. We consider the values
\begin{equation}
	A \in \left\{0,0.15, 0.25,0.5,1,1.5,2.5,4,\infty\right\}.
\end{equation}
The value $A=0$ corresponds to a Neumann boundary conditions, while $A>0$ imposes mixed boundary conditions. Dirichlet boundary conditions, corresponding to $A\to \infty$ in the continuum theory, are obtained by setting $Q=0$ in equation~\eqref{eq:coupling matrix}.

The total entanglement entropy is given by~\eqref{eq:SEE_total}. For each value of $N$ and $n$, we calculate the contribution of the $\ell$-sectors from $\ell=0$ to $\ell=3\cdot 10^4$, and use an extrapolation in order to approximate the value of the infinite sum. Following \cite{Boutivas:2024lts,Boutivas:2025ksp}, we calculate the truncated sum
\begin{equation}
	S_{\textrm{EE}}(n,N,A;\ell_{\textrm{max}})=\sum_{\ell=0}^{\ell_{\textrm{max}}}(2\ell+1)S_{\ell m}^{\textrm{EE}}(n,N,A)
\end{equation}
for various values of $\ell_\text{max}$ and study its behavior for large values of $\ell_{\textrm{max}}$. We find that it can be fitted by the
expression
\begin{equation}
	S_{\textrm{EE}}(n,N,A;\ell_{\textrm{max}})	 = S_{\infty}(n,N,A)
	+ \sum_{i=1}^{i_\textrm{max}}\frac{1}{\ell_\textrm{max}^{2i}}\left(
	a_i(n,N,A)+b_i(n,N,A)\ln \ell_\text{max}\right).
\end{equation}
By including a sufficient number of subleading terms, we can extrapolate to the infinite-$\ell_\text{max}$ limit 
\begin{equation}
	S_\infty(n,N,A)=\lim_{\ell_\textrm{max}\rightarrow\infty}S_{\textrm{EE}}(n,N,A;\ell_\textrm{max})
\end{equation}
with high accuracy.

Next, $S_{\infty}(n,N,A)$ is studied as a function of the ratio ${a}/{\epsilon}=({2}/{\pi})(N+1)$. It can be expanded as
\begin{equation}
	S_{\infty}(n,N,A)= \frac{a^2}{\epsilon^2} S^{(2)}(n,A)+ S_{\log}^{(0)}(n,A)\ln\frac{a}{\epsilon}+ S^{(0)}(n,A)+R(n,N,A),
	\label{eq:UV_expansion}
\end{equation}
where the remainder $R(n,N,A)$ is
\begin{equation}
	R(n,N,A)=\sum_{i=1}^{i_\textrm{max}} \frac{\epsilon^i}{a^i}S^{(-i)}(n,A)+\ln\frac{a}{\epsilon}\sum_{j=1}^{j_\textrm{max}} \frac{\epsilon^j}{a^j}S_{\log}^{(-j)}(n,A).
\end{equation}
Even though the remainder vanishes for $\epsilon\rightarrow0$, its inclusion  is necessary in order to calculate accurately the terms $S^{(2)}(n,A)$, $S_{\log}^{(0)}(n,A)$ and $S^{(0)}(n,A)$. These are computed with high precision, and their behavior, as functions of the position of the entangling surface $w_R$ 
\begin{equation}
 	\frac{w_R}{a}=\frac{\pi}{2}\frac{n+\frac{1}{2}}{N+1},
 	\label{eq:wr}
\end{equation}
is examined below. 
 
Custom C++ code relying on the package Eigen for linear algebra was used for the numerical calculations. The code implements 128-bit precision arithmetic, corresponding to 33--35 significant digits. 

\subsection{Results}
\label{sec:results}

The numerical analysis of the data shows that the dominant divergent term is of the form
\begin{equation}
	S^{(2)}(n)=  d_1\sin^2\frac{w_R}{a},
	\label{eq:square_term}
\end{equation}
for all values of $A$ that determine the boundary conditions. The coefficient $d_1$ can be computed with high accuracy. It takes the value
\begin{equation}
	d_1\simeq 0.29543145
	\label{eq:d1_value}
\end{equation}
for all values of $A$. This result is in complete agreement with \cite{Boutivas:2025ksp}. For $w_R \ll a$ the curvature of the AdS has a negligible effect, and the above expression and the numerical value of $d_1$ reproduce correctly the leading UV divergence of the entanglement entropy in flat space \cite{Srednicki:1993im}, as expected. For large $w_R/a$ there seems to be a discrepancy in characterizing the above expression as an area term, because the proper area of the entangling surface is $\Ac=4\pi a^2\tan^2{\frac{w_R}{a}}$. The difference can be traced to the regularization based on the discretization of the radial coordinate $w/a$, which differs drastically from the regularization through the discretization of the coordinate $r=a\tan({w}/{a})$ that takes values between 0 and infinity. As discussed in \cite{Boutivas:2025ksp}, there is a relative factor of $1/\cos^2{\frac{w}{a}}$ for the number of degrees of freedom in the discretization of the radial coordinate $r$ in~\eqref{eq:globalr} relative to the discretization of the tortoise coordinate $w$ in~\eqref{eq:globalw}. When this is taken into account, the expected dependence on the proper area is reproduced.
 
\begin{figure}[t]
	\centering
%	\vspace{-1cm}
	\begin{picture}(95,57)
		\put(5,1){\includegraphics[angle=0,width=0.9\textwidth]{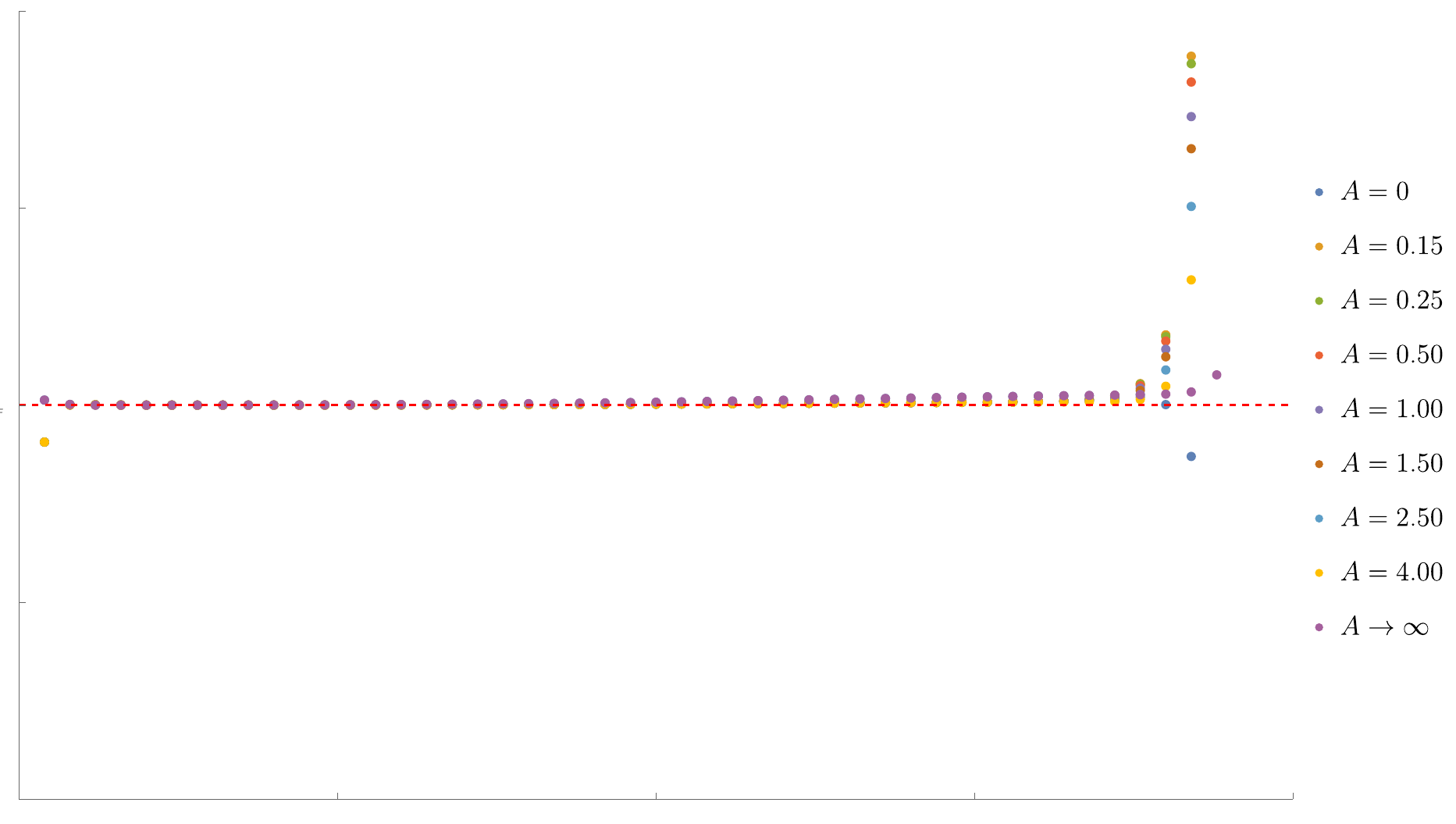}}
		\put(0,53){{\Large $-90\,S_{\log}^{(0)}$}}
		\put(86.3,1.9){{\Large $\frac{w_R}{a}$}}
		\put(44.8,0){{\small $\frac{\pi}{4}$}}
		\put(25,0){{\small $\frac{\pi}{8}$}}
		\put(64,0){{\small $\frac{3\pi}{8}$}}
		\put(84.25,0){{\small $\frac{\pi}{2}$}}
		\put(2.7,1.8){{\small $0.0$}}
		\put(2.7,14.0){{\small $0.5$}}
		\put(2.7,26.1){{\small $1.0$}}
		\put(2.7,38.4){{\small $1.5$}}
		\put(2.7,50.5){{\small $2.0$}}		
	\end{picture}
	\caption{Numerical fits for the determination of $S_{\log}^{(0)}$ as a function of $w_R/a$ for various values of $A$. The data indicate with high precision that $S_{\log}^{(0)}=-{1}/{90}$.}
	\label{fig:plot2}
%\end{figure}
%
%\begin{figure}[t]
	\centering
%	\vspace{-1cm}
	\begin{picture}(95,57)
		\put(5.625,1.25){\includegraphics[angle=0,width=0.87375\textwidth]{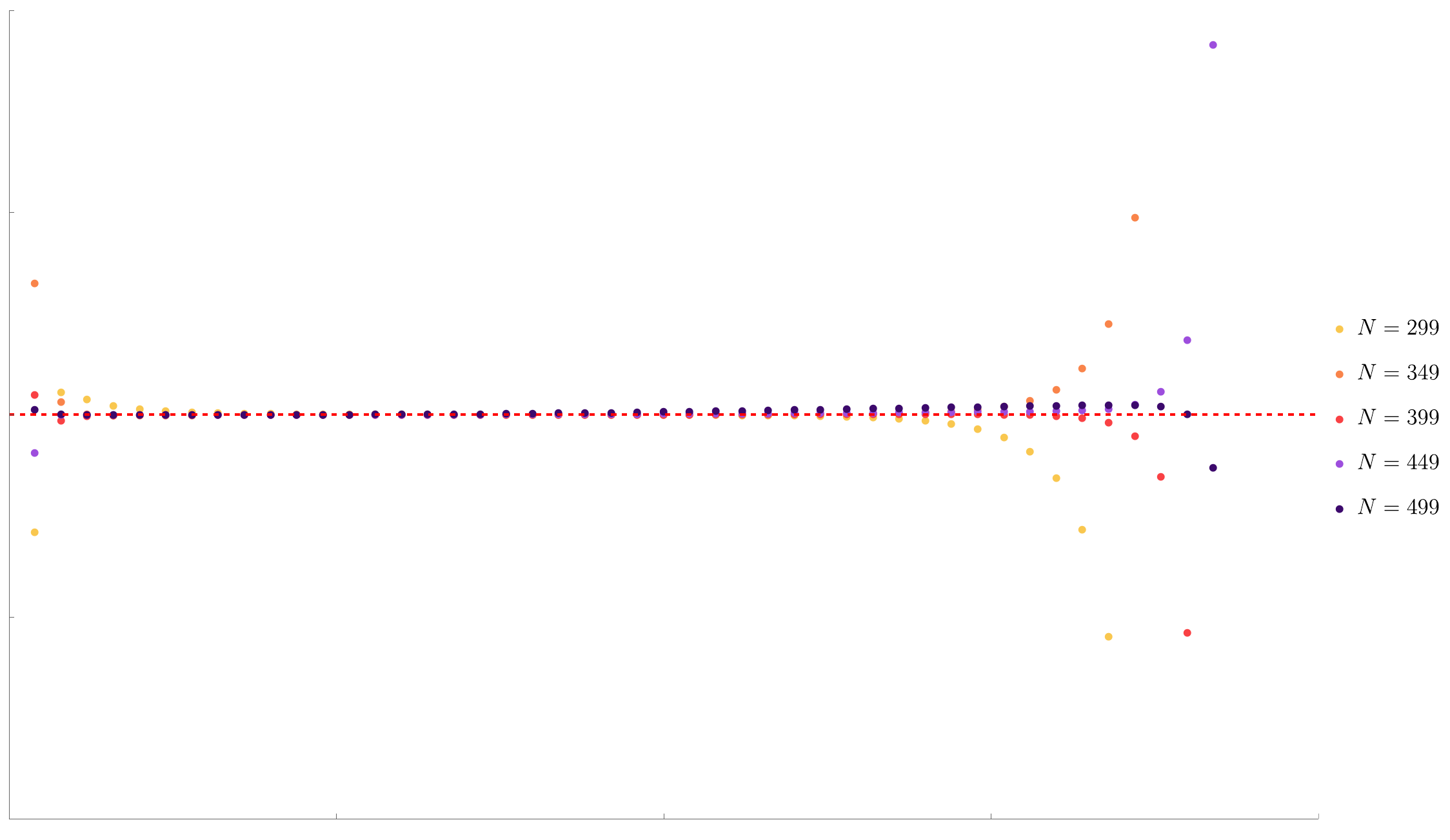}}
		\put(0,53){{\Large $-90\,S_{\log}^{(0)}$}}
		\put(86.3,1.9){{\Large $\frac{w_R}{a}$}}
		\put(44.8,0){{\small $\frac{\pi}{4}$}}
		\put(25,0){{\small $\frac{\pi}{8}$}}
		\put(64,0){{\small $\frac{3\pi}{8}$}}
		\put(84.25,0){{\small $\frac{\pi}{2}$}}
		\put(2.7,1.8){{\small $0.0$}}
		\put(2.7,14.0){{\small $0.5$}}
		\put(2.7,26.1){{\small $1.0$}}
		\put(2.7,38.4){{\small $1.5$}}
		\put(2.7,50.5){{\small $2.0$}}
	\end{picture}
	\caption{Convergence of the numerical determination of $S^{(0)}_{\log}$ as a function of $w_R/a$ for Neumann boundary conditions ($A=0$). The datasets correspond to $N=299,349,399,449$ and $499$. Increasing $N$ enlarges the range of data available for fitting the asymptotic expansion in powers of $\epsilon$, thereby improving the extraction of the coefficient of the $\ln(a/\epsilon)$ term, identified with $S_{\log}^{(0)}$. The results converge towards $S_{\log}^{(0)}=-1/90$.}
	\label{fig:plot2convergence}
\end{figure}

The coefficient of the divergent term proportional to $\ln\epsilon$ is depicted in Fig.~\ref{fig:plot2} for various values of $A$. For almost the entire range of $w$, it takes the universal value $S_{\log}^{(0)}(n,A)=-{1}/{90}$ for all choices of boundary conditions. Deviations from this value are visible close to the endpoints, namely near $w_R/a\simeq 0$ and near the conformal boundary $w_R/a\simeq \pi/2$. At first sight, one might suspect that these deviations reflect a genuine boundary-condition effect, especially near the conformal boundary, whose influence is expected to be more pronounced. However, this interpretation is not supported by our numerical checks. By using more refined lattices through the increase of the number of lattice points $N$, the fits are improved visibly, as shown in Fig.~\ref{fig:plot2convergence}, and the different divergent terms in the cutoff expansion of the entropy~\eqref{eq:UV_expansion} are isolated with higher accuracy.
This improved extraction increases the computational time substantially, but demonstrates that the endpoint deviations are reduced with increasing $N$. We therefore interpret these departures from $-1/90$ as numerical artefacts, rather than physical effects of the boundary conditions. It must also be kept in mind that the entropy of the finite, discretized system must drop to zero when the entangling surface approaches the boundary, as the subsystem then becomes the whole system. 
% This effect is caused by the finite number of degrees of freedom allowed by the discretization. 
For large but finite $N$, the entropy will make a sharp turn towards zero very close to the boundary \cite{Boutivas:2025ksp}. The numerical analysis suggests that the deviations from a constant value observed near the boundary in Fig.~\ref{fig:plot2} are generated by this effect. On the other hand, in the continuum limit there is always an infinite number of degrees of freedom between the last entangling surface and the boundary, so that the entropy will continue to grow as a function of $w_R$. Furthermore, it must be noted that genuine effects associated with the AdS length, such as a logarithmically divergent term for a theory away from the conformal point, which are studied in detail in \cite{Boutivas:2025ksp}, actually become visible for much smaller values of the ratio $w/a$.

Before proceeding to the discussion of the UV-finite part of the entanglement entropy, we must address a conceptual issue that arises because of the UV-cutoff dependence. As discussed above, the regularization scheme affects drastically the form of the leading UV-divergent term. On the other hand, the coefficient of the logarithmically divergent term is not affected and obtains a universal value related to the $A$-type conformal anomaly of the theory. However, the arbitrariness in defining the UV cutoff generates an ambiguity that affects the UV-finite part. In general terms, if the UV cutoff is redefined through the multiplication by a $w$-dependent function $f(w)$, as was done for the leading term, a finite contribution $\sim \ln f(w_R)$ to the entropy is automatically induced. It is, therefore, difficult to assign physical significance to the finite part, unless the UV cutoff is given a physical meaning. This requires that gravity be taken into account beyond the level of an external background. 

For the problem at hand, the ambiguity is related to the difference in the discretization of the radial coordinate $r$ in~\eqref{eq:globalr} relative to the discretization of the tortoise coordinate $w$ in~\eqref{eq:globalw}. Fortunately, both regularizations coincide near the origin (i.e., for $w\to 0$), where they also approach the flat-space limit. This raises the expectation that the finite part of the entropy for small $w$ can be used in order to extract unambiguous information about the theory via a comparison to the flat space theory. In particular, the dependence on the boundary conditions at the conformal boundary can be investigated reliably. In fact, it is a very interesting question whether boundary conditions imposed for $w/a\to  {\pi}/{2}$, or $r/a \to \infty$, can affect the entanglement entropy for $w/a\simeq r/a \ll 1$. We focus on this regime in the following.

The corresponding flat-space result \cite{Srednicki:1993im, Solodukhin:2008dh,Casini:2009sr,Lohmayer:2009sq} indicates that we should expect the entanglement entropy to contain a UV-divergent logarithmic contribution $-(1/90)\log(R/\epsilon)$, with $R=a\tan(w_R/a)$, whose coefficient is associated with the conformal $A$-type anomaly. In the present setup, this term is split between the logarithmically divergent part $S^{(0)}_l$ and the UV-finite part $S^{(0)}$ of the entanglement entropy. After the cutoff-dependent contribution $-(1/90)\log(a/\epsilon)$ has been identified, the remaining $R$-dependent part must arise from a contribution $-(1/90)\log (R/a)$ contained in the constant term $S^{(0)}$. (We have used the AdS length $a$ for normalization.) It is not straightforward to isolate this logarithmic dependence directly from $S^{(0)}$, since the latter may also contain additional finite contributions, including terms proportional to $R^2$, whose coefficients can depend on the boundary-condition parameter $A$. For this reason, we act on $S^{(0)}$ with the differential operator
\begin{equation}
\mathcal{D}_R
=
45
\left(
R^2\frac{d^2}{dR^2}
-R\frac{d}{dR}
\right),
\end{equation}
which is chosen so that $\mathcal{D}_R[-(1/90)\log R]=1$, while it annihilates a pure $R^2$ contribution. It therefore provides a useful diagnostic for the presence of the universal logarithmic term in the finite part of the entropy. This operator was used in \cite{Abate:2024nyh,Abate:2026apg} in order to define an $a$-function in $3+1$ dimensions through the entanglement entropy. As the $a$-function must reproduce the coefficient of the conformal anomaly at the fixed points of the renormalization-group flow, it is the appropriate tool for identifying the logarithmic term in our framework as well.

\begin{figure}[t]
	\centering
%	\vspace{-1cm}
	\begin{picture}(93,55)
		\put(3,1){\includegraphics[angle=0,width=0.9\textwidth]{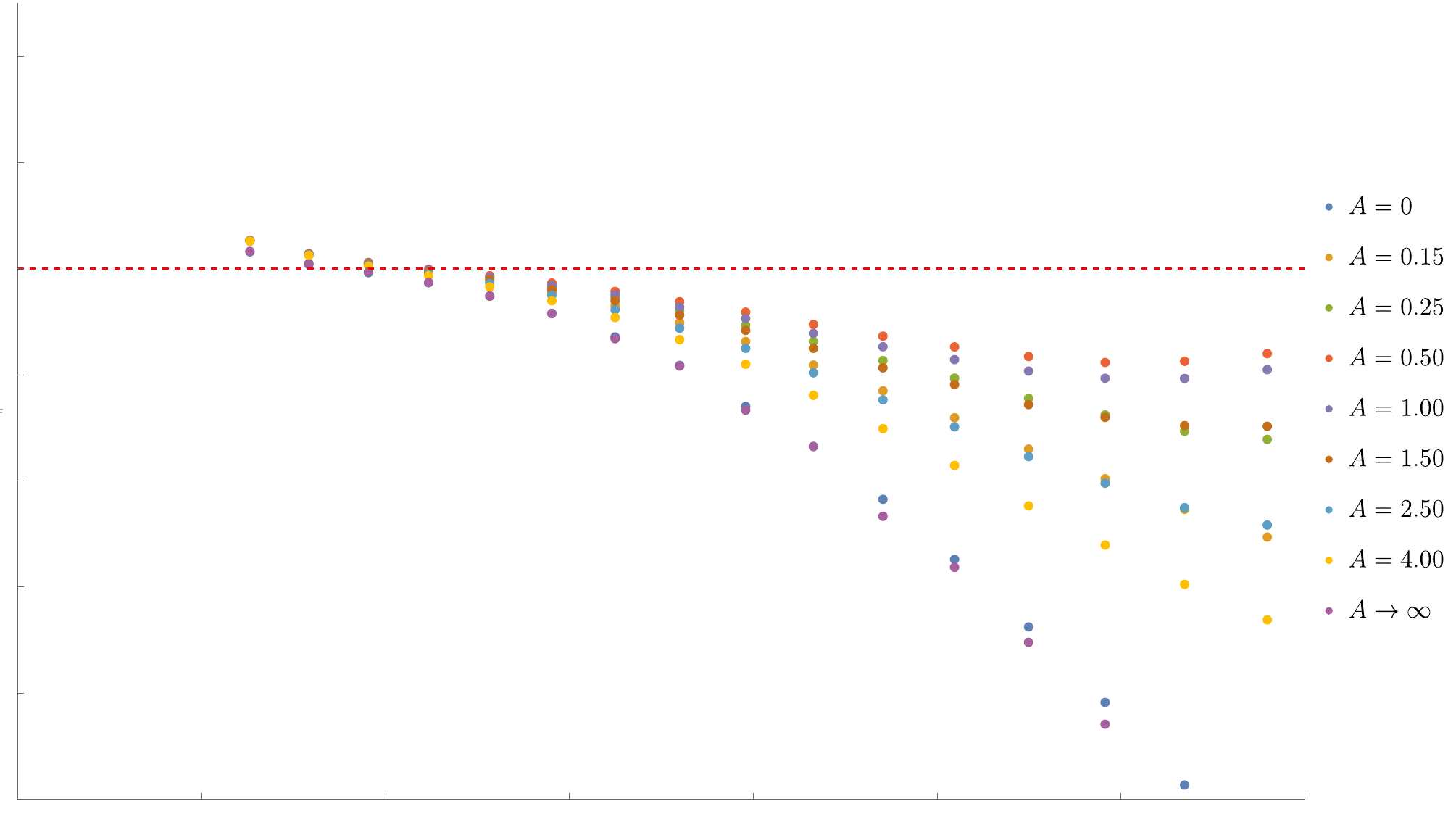}}	
		\put(0,52){{$\mathcal{D}_R S^0(R)$}}
		\put(85.5,1.625){{\Large $\frac{R}{a}$}}
		\put(14,0){{\small $0.1$}}		
		\put(25.375,0){{\small $0.2$}}
		\put(36.75,0){{\small $0.3$}}
		\put(48.125,0){{\small $0.4$}}
		\put(59.5,0){{\small $0.5$}}
		\put(70.875,0){{\small $0.6$}}
		\put(82.26,0){{\small $0.7$}}
		\put(0.7,1.8){{\small $0.0$}}
		\put(0.7,8.3){{\small $0.2$}}
		\put(0.7,14.8){{\small $0.4$}}
		\put(0.7,21.5){{\small $0.6$}}
		\put(0.7,27.9){{\small $0.8$}}
		\put(0.7,34.5){{\small $1.0$}}
		\put(0.7,41.2){{\small $1.2$}}
		\put(0.7,47.6){{\small $1.4$}}	
	\end{picture}
	\caption{Numerical evaluation of $\mathcal{D}_R S^{(0)}(R)$, where $\mathcal{D}_R = 45\left(R^2 d^2/dR^2 - R d/dR\right)$ and $R=a\tan(w_R/a)$, for several values of $A$. The dashed red line denotes $\mathcal{D}_R S^{(0)}(R)=1$. The convergence of the data to this value in the small-$R$ regime provides numerical evidence for the existence of a term $ -(1/90)\log R$ in $S^{(0)}(R)$. Combined with the cutoff-dependent contribution $-(1/90)\log(1/\epsilon)$, this yields the universal logarithmic term $-(1/90)\log(R/\epsilon)$, also present in flat space.}
	\label{fig:plot3}
\end{figure}
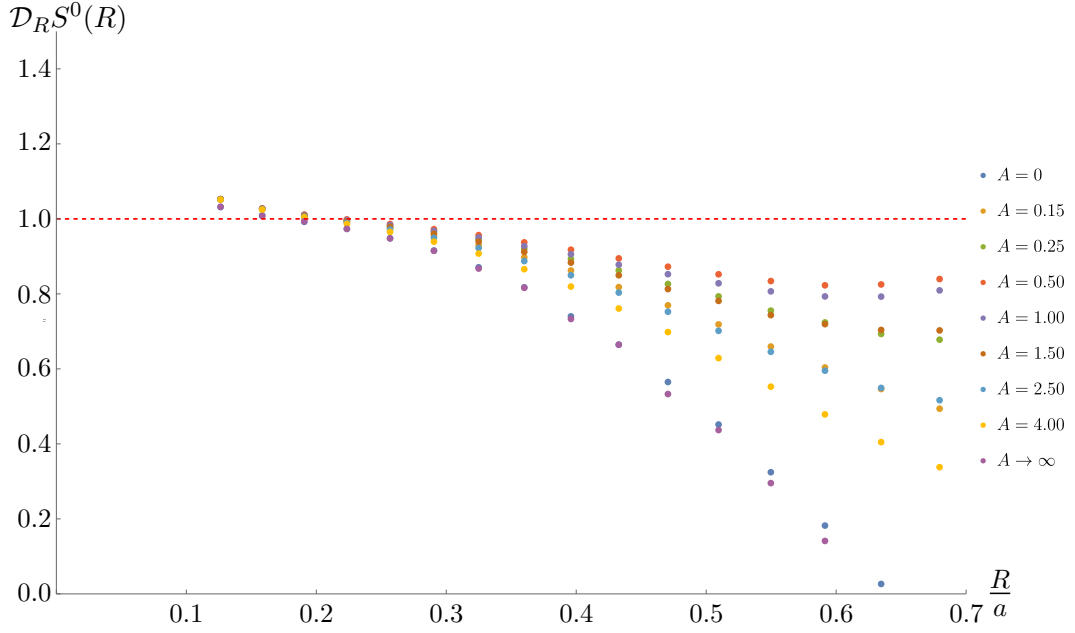

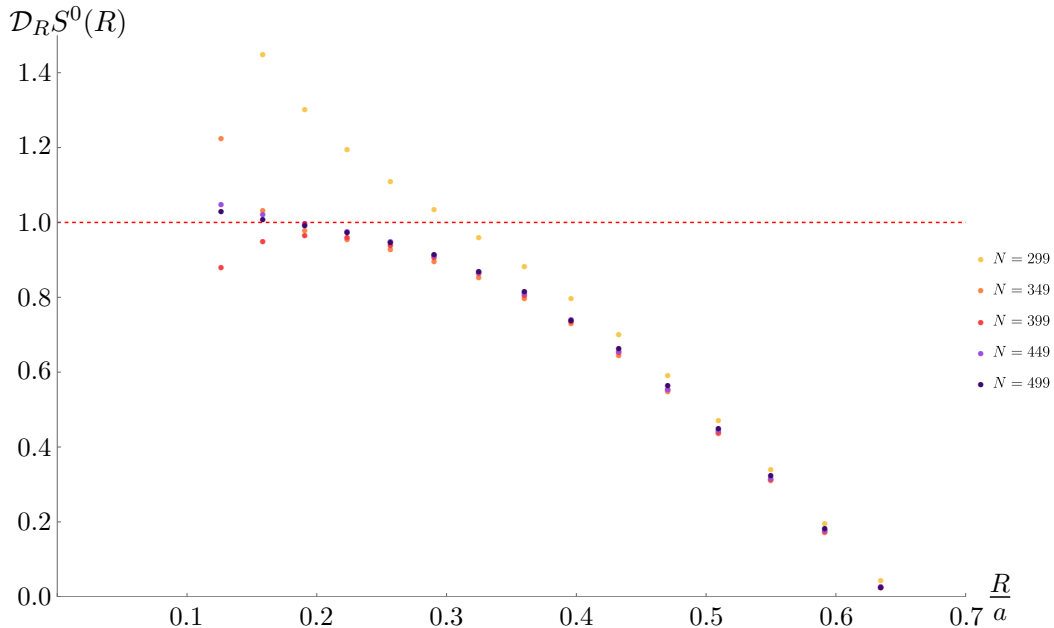
\begin{figure}[t]
	\centering
%	\vspace{-1cm}
	\begin{picture}(93,55)
		\put(3.625,1.25){\includegraphics[angle=0,width=0.88\textwidth]{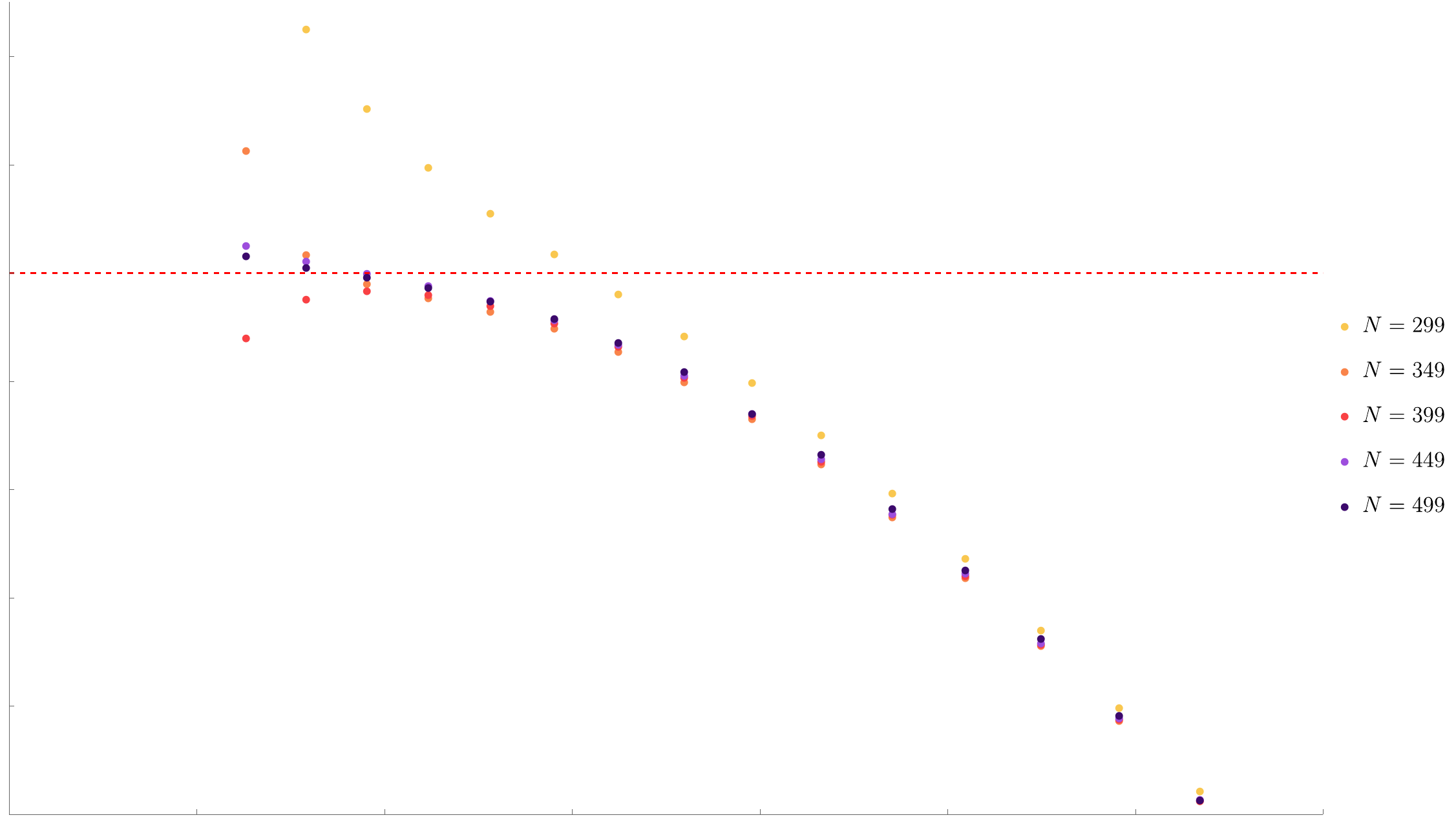}}	
		\put(0,52){{$\mathcal{D}_R S^0(R)$}}
		\put(85.5,1.625){{\Large $\frac{R}{a}$}}
		\put(14,0){{\small $0.1$}}		
		\put(25.375,0){{\small $0.2$}}
		\put(36.75,0){{\small $0.3$}}
		\put(48.125,0){{\small $0.4$}}
		\put(59.5,0){{\small $0.5$}}
		\put(70.875,0){{\small $0.6$}}
		\put(82.26,0){{\small $0.7$}}
		\put(0.7,1.8){{\small $0.0$}}
		\put(0.7,8.3){{\small $0.2$}}
		\put(0.7,14.8){{\small $0.4$}}
		\put(0.7,21.5){{\small $0.6$}}
		\put(0.7,27.9){{\small $0.8$}}
		\put(0.7,34.5){{\small $1.0$}}
		\put(0.7,41.2){{\small $1.2$}}
		\put(0.7,47.6){{\small $1.4$}}
	\end{picture}
\caption{Convergence of the numerical determination of $\mathcal{D}_R S^{(0)}(R)$ as a function of $R/a$ for Neumann boundary conditions. The datasets correspond to $N=299,349,399,449$ and $499$. Increasing $N$ enlarges the range of data available for fitting the asymptotic expansion in powers of $\epsilon$, thereby improving the extraction of  $S^{(0)}$. The results converge towards $\mathcal{D}_R S^{(0)}(R)=1$ in the small-$R$ regime.}
	\label{fig:plot3convergence}
\end{figure}

The result of the calculation is shown in Fig.~\ref{fig:plot3}. For all values of $A$, the data approach the value $\mathcal{D}_R S^{(0)}=1$ in the small-$R$ regime, providing numerical evidence for the existence of the term $-(1/90)\log R$ in $S^{(0)}$. Small deviations from this value, observable in Fig.~\ref{fig:plot3}, decrease with $N$ and tend to vanish at the continuous limit, as shown in Fig.~\ref{fig:plot3convergence}. Combined with the cutoff-dependent contribution, this reconstructs the expected universal term $-(1/90)\log(R/\epsilon)$. At the same time, the curves exhibit clear $A$-dependent deviations away from the strictly small-$R$ limit. These show that, although the coefficient of the logarithmic term is universal, the finite part of the entropy does receive  corrections that depend on the boundary conditions, visible even for small values of $R$.

It becomes apparent from the above that, in order to isolate the subleading contribution to the finite part of the entropy that is sensitive to the boundary conditions, one can subtract the Dirichlet result from the one for an arbitrary $A$ that determines specific boundary conditions. Therefore, we define 
\begin{equation}
	\Delta S^{(0)}(A;R)
	:=
	S_A^{(0)}(R)-S_{A\to\infty}^{(0)}(R)\, ,
	\label{eq:DeltaS_definition}
\end{equation}
where $A\to\infty$ corresponds to Dirichlet boundary conditions. This subtraction removes the part common to all boundary conditions and isolates the UV-finite contribution that depends explicitly on them. The UV-finite term $S_A^{(0)}(R)$ is not universal by itself, since a redefinition of the UV cutoff may shift it by finite contributions inherited from the area-law and logarithmic terms. However, the coefficients of these UV-divergent terms have been shown above to be independent of the boundary conditions. The corresponding scheme-dependent shifts therefore cancel in $\Delta S^{(0)}(A;R)$, making this difference independent of the UV regularization scheme, provided that the same prescription is used for all boundary conditions.

Before presenting the results, it is instructive to recall some known facts about the presence of IR contributions to the entanglement entropy. Corresponding terms in the entropy have been identified in dS space \cite{Boutivas:2024sat,Boutivas:2024lts} and for the Einstein universe \cite{Boutivas:2025rdf}. It has been shown that such terms arise from the $\ell=0$ sector of the theory. This has motivated us to examine separately the contributions from this sector, for which an analytical treatment is often feasible. In appendix~\ref{appendix} we consider the $(1+1)$-dimensional scalar theory describing the $\ell=0$ sector, on a segment of length $L/a=\pi/2$. In order to mimic the $(3+1)$-dimensional case, we impose Dirichlet boundary conditions on the left boundary and Neumann on the right. For small $R$, a fully analytical calculation is feasible. The end result for $A=0$ that corresponds to Neumann boundary conditions on the right boundary is
\begin{equation} 
	\Delta S^{(0)}(0;R)=\frac{1}{6}\frac{R^2}{a^2}.
	\label{eq:1p1A0}
\end{equation}
A derivation of a similar solution for mixed boundary conditions is impeded by the fact that the spectrum of eigenvalues of the \Schrodinger Hamiltonian is not regular in the general case, as it is determined by the transcendental equation~\eqref{eq:gbc}. However, a numerical calculation can be performed with high accuracy.

\begin{figure}[t]
	\centering
%	\vspace{-1cm}
	\begin{picture}(95,55)
		\put(5,0){\includegraphics[angle=0,width=0.9\textwidth]{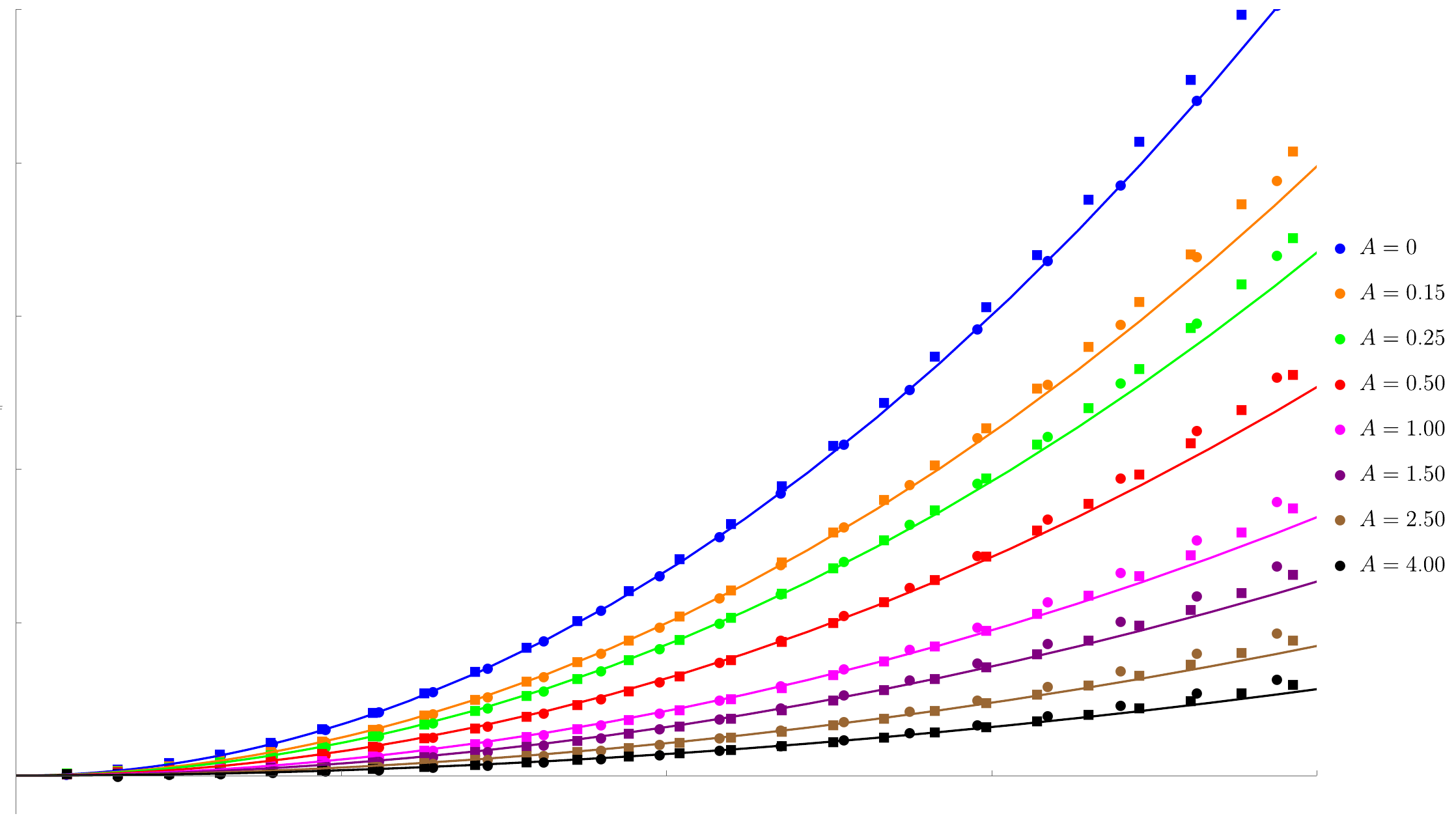}}
		\put(0.5,52){{$ \Delta S^0(A;R)$}}
		\put(88.3,1.75){{\Large $\frac{R}{a}$}}			
		\put(24.7,0){{\small $0.2$}}		
		\put(44.8,0){{\small $0.4$}}
		\put(64.9,0){{\small $0.6$}}
		\put(84.9,0){{\small $0.8$}}
		\put(1.4,1.8){{\small $0.00$}}
		\put(1.4,11.3){{\small $0.02$}}
		\put(1.4,20.7){{\small $0.04$}}
		\put(1.4,30.2){{\small $0.06$}}
		\put(1.4,39.5){{\small $0.08$}}
		\put(1.4,49){{\small $0.10$}}
	\end{picture}
	\caption{The difference $\Delta S^{(0)}(A;R)=S^{(0)}_A(R)-S^{(0)}_{A\rightarrow\infty}(R)$ between the UV-finite term of the entanglement entropy for a general boundary-condition parameter $A$ and the Dirichlet result obtained for  $A\rightarrow\infty$, plotted as a function of $R=\tan(w_R/a)$. For each value of $A$, circles denote the $(3+1)$-dimensional data, while squares denote the corresponding $(1+1)$-dimensional data, arising from the $\ell=0$ sector. The solid lines are quadratic fits to the $(3+1)$-dimensional data of the form $\Delta S^{(0)}(A;R)=c_A (R/a)^2$, quantifying the leading small-$R$ dependence.
		% of the effect of
		%the  boundary conditions.
}
	\label{fig:plot4}
\end{figure}
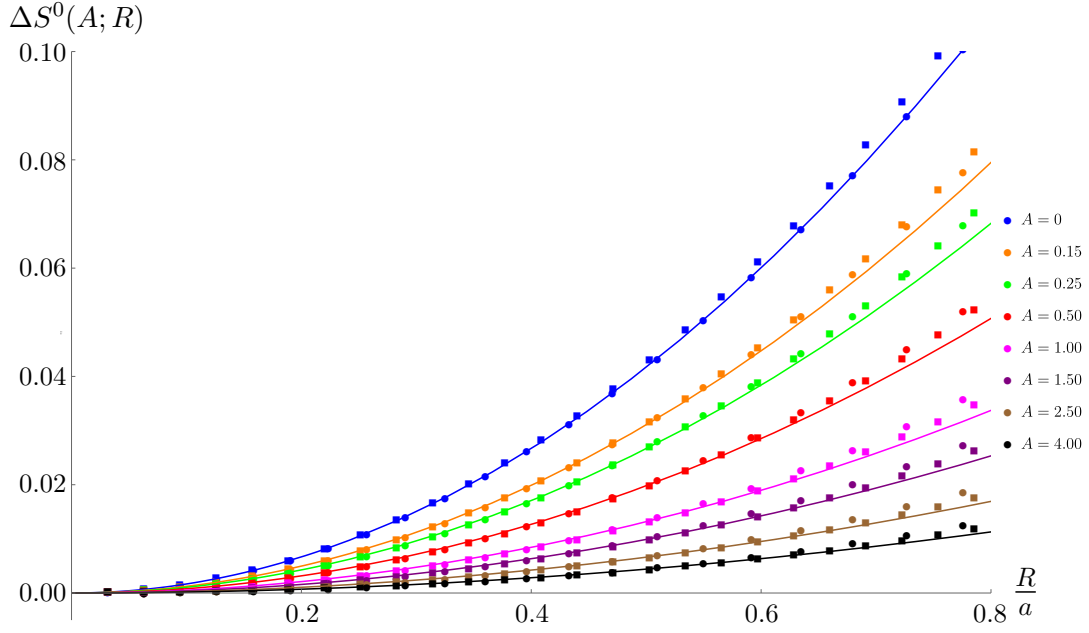

The quantity $\Delta S^{(0)}(A;R)$ is shown in Fig.~\ref{fig:plot4}, where circles denote the full $(3+1)$-dimensional data, while squares denote the corresponding $\ell=0$ contribution. The close agreement between the two sets of points shows that the  part of the finite entropy that depends on the boundary conditions is dominated by the $\ell=0$ sector. For small $R$, all curves are well described by a quadratic function
\begin{equation}
	\Delta S^{(0)}(A;R)=c_A \frac{R^2}{a^2},
	\label{eq:1p1A}
\end{equation}
represented by the solid lines. The agreement is particularly robust for the Neumann case, where the quadratic fit remains accurate over a wider range of $R$. Deviations from the purely quadratic behavior appear at large $R$, indicating the presence of additional subleading corrections.

\begin{figure}[t]
	\centering
%	\vspace{-1cm}
	\begin{picture}(86,55)
		\put(3,0){\includegraphics[angle=0,width=0.8\textwidth]{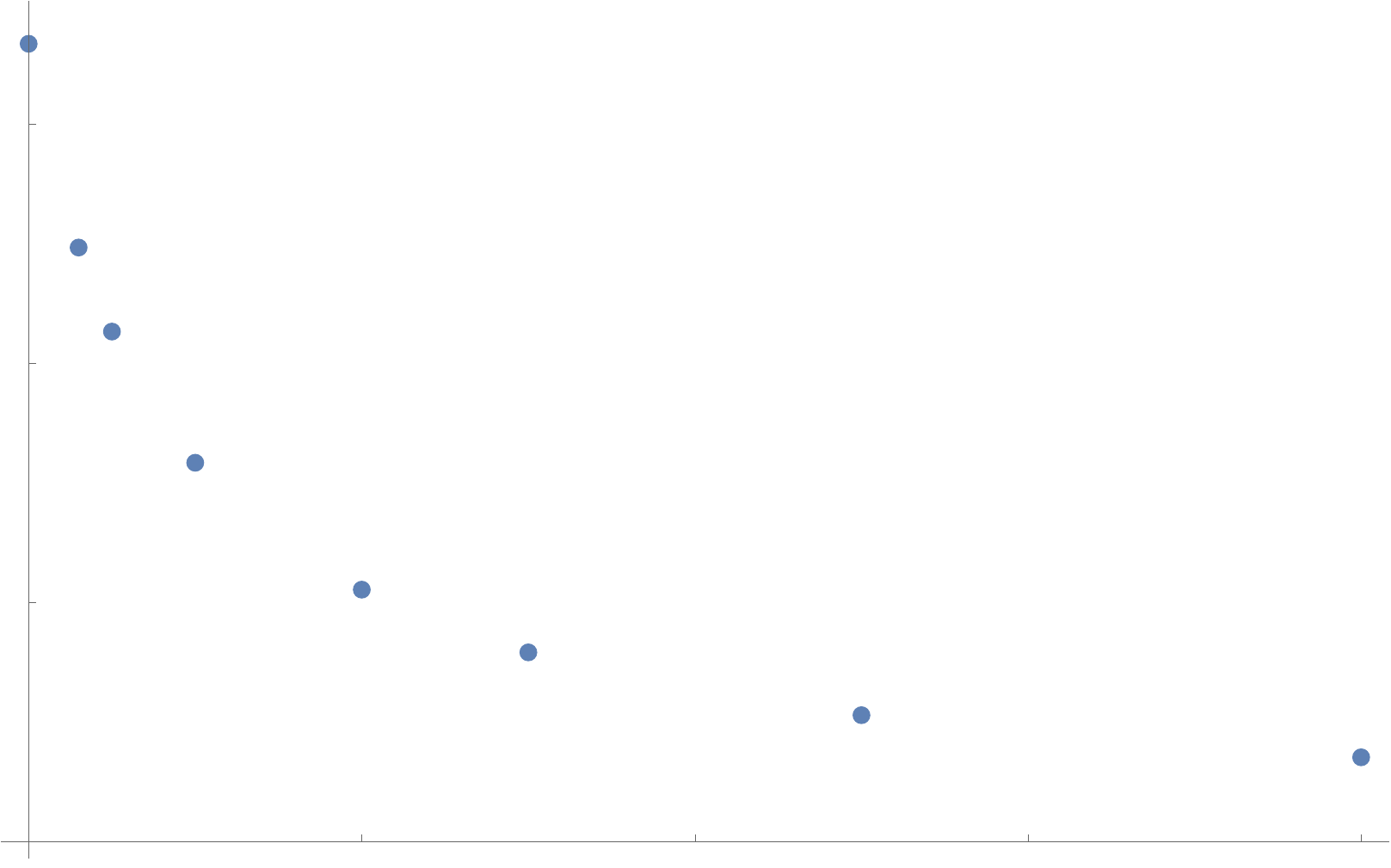}}
		\put(3,51.3){{\Large $c_A$}}
		\put(83.5,0.3){{\Large $A$}}
		\put(23.3,-1){{\small $1$}}		
		\put(42.375,-1){{\small $2$}}
		\put(61.65,-1){{\small $3$}}
		\put(80.75,-1){{\small $4$}}
%		\put(0.6,0.8){{\small $0.00$}}
		\put(0,14.4){{\small $0.05$}}
		\put(0,28.1){{\small $0.10$}}
		\put(0,42.0){{\small $0.15$}}	
	\end{picture}
	\caption{The coefficient $c_A$ extracted from the small-$R$ fits $\Delta S^{(0)}(A;R)=c_A R^2$, as a function of the boundary-condition parameter $A$. 
		%The limited number of data points does not allow a unique determination of the functional dependence on $A$, but the behaviour is well described by a decreasing hyperbola. For $A=0$, corresponding to the Neumann boundary condition, we find $c_A=1/5.997\simeq 1/6$, in agreement with the analytical calculations for the $(1+1)-$dimensional problem, while $c_A$ decreases towards the Dirichlet limit at large $A$.
		}
	\label{fig:plot5}
\end{figure}
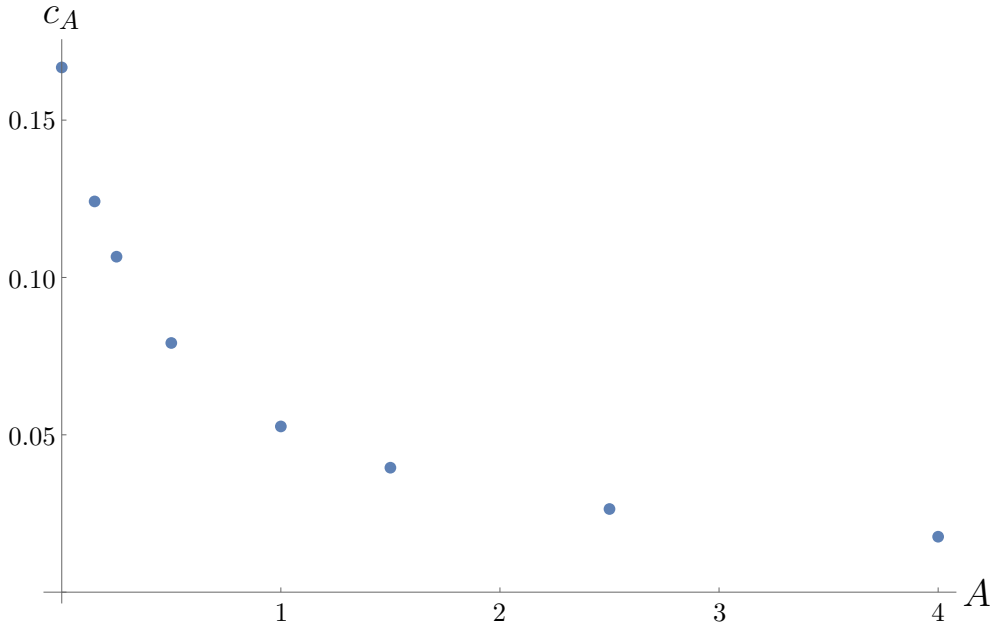

The coefficient $c_A$ decreases as $A$ increases, consistently with the approach to the Dirichlet limit, where $\Delta S^{(0)}$ vanishes by construction. The values of $c_A$ extracted from the quadratic fits of Fig.~\ref{fig:plot4} are shown in Fig.~\ref{fig:plot5} as a function of the boundary-condition parameter $A$. The coefficient is maximal for $A=0$, corresponding to Neumann boundary conditions, where it takes the value $c_A=1/6$ derived analytically in appendix~\ref{appendix}. It decreases monotonically as $A$ is increased, following a power law for large $A$. Because of our lack of analytical understanding of the smooth behavior depicted in in Fig.~\ref{fig:plot5}, we do not present any fits, as these would be rather arbitrary.

\section{Summary and Discussion}
\label{sec:discussion}

In this work we continued the analysis of \cite{Boutivas:2025ksp} by focusing on the role of boundary conditions in the calculation of the entanglement entropy of a scalar field in AdS space. For a range of negative values of the squared mass, there is more than one normalizable field mode, with distinct physical interpretation. The modes are characterized by their asymptotic form at large distances. Selecting a particular mode can be achieved by imposing appropriate boundary conditions at the conformal boundary that lead to self-adjoint extensions of the effective \Schrodinger Hamiltonian. In general, implementing these conditions in a numerical study is difficult, as one must fix the asymptotic form of the eigenmodes of the discretized system, and not just their value at a particular point. Fortunately, the conformal point provides a particularly convenient realization, for which the boundary conditions simplify: they can be fixed through the ratio of the derivative of the field with respect to the tortoise coordinate~\eqref{eq:tortoise} to its value at the conformal boundary. We focused on this case, which offers the additional advantage that additional scales, such as an explicit mass term, that complicate the picture and have been considered already in \cite{Boutivas:2025ksp}, are absent. The parameter $A$, defined in equation (\ref{eq:gbc}), allows us to select various mixed boundary conditions, with the limits $A\to 0$ and $A\to \infty$ corresponding to Neumann and Dirichlet, respectively. 

The expected form of the entanglement entropy for a spherical entangling surface in $3+1$ dimensions is given by~\eqref{eq:SEE_expansion_mald}. For a conformally coupled theory with  Dirichlet boundary conditions, the values of the parameters $d_3$ and $d_4$ lead to the cancellation of the corresponding terms. However, allowing for mixed boundary conditions introduces a new energy scale, associated with the ratio of the field to its derivative. Measuring this scale in units of the AdS scale allows the definition of a new dimensionless quantity such as the ratio $A$ defined in equation~\eqref{eq:gbc}. It is then possible that a new logarithmic divergence may arise, in complete analogy to the terms proportional to $d_3$ and $d_4$ in (\ref{eq:SEE_expansion_mald}). Our analysis has revealed no such term (see Fig.~\ref{fig:plot2}), consistently with the general expectation that long-distance boundary effects should not alter the UV structure of physical quantities.

Our study then moved to the UV-finite part of the entanglement entropy. The level of numerical accuracy that we achieved is sufficient in order to analyze this part. However, conceptual issues arise, related to the contamination of the finite part by ambiguities originating in the choice of the UV cutoff, as we discussed in the previous section. For this reason, we focused on the range of small entangling radii $R$, for which these ambiguities are resolved through comparison with the flat-space limit. For small $R$, the leading UV-finite term is the one proportional to  $d_6$ in~\eqref{eq:SEE_expansion_mald}. Our analysis showed that this term is universal, independently of the boundary conditions (see Fig.~\ref{fig:plot3}), and its coefficient takes the appropriate value $d_6=-1/90$ in order to combine with the UV-divergent term with $d_2=-1/90$ so that the known flat-space result is reproduced. 

We next considered the subleading finite term for small $R$, which is the one proportional to the coefficient $d_5$ in  (\ref{eq:SEE_expansion_mald}). A dependence on the boundary conditions was seen clearly in the numerical analysis, in which we calculated the difference of the finite part of the entanglement entropy between general mixed boundary conditions for various values of $A$ and the case of Dirichlet boundary conditions with $A\to \infty$. Numerical fits of the results demonstrate the quadratic dependence of the entropy on $R$ (see Fig.~\ref{fig:plot4}). The coefficient $c_A$ of the quadratic term, defined in equations~\eqref{eq:DeltaS_definition} and ~\eqref{eq:1p1A}, interpolates smoothly between $1/6$ and $0$ when $A$ changes from 0 (Neumann) to $\infty$ (Dirichlet). 

Our numerical calculations are supported by analytical means. As is apparent from Fig.~\ref{fig:plot4}, the UV-finite quadratic contribution to the entropy for small $R$ is linked to the $\ell=0$ angular-momentum sector. This amounts to a $(1+1)$-dimensional massless theory on a flat background, with Dirichlet boundary conditions at one end of the total system and mixed boundary conditions at the other, depending on the value of $A$. It is possible to obtain exact analytical results for the entropy for $A=0$ (Neumann) and $A=\infty$ (Dirichlet), as the eigenvalues follow regular patterns in these cases that make the calculation feasible. The results for an infinite total system, or for entangling radii much smaller than the size of a finite system, are given by equations~\eqref{eq:entropy1ddn} and~\eqref{eq:entropy1ddd}. The difference of the two expressions explains the value $c_A=1/6$ for $A=0$ observed in Fig.~\ref{fig:plot5}.

Our results reveal the effect of the boundary conditions on the form of the entanglement entropy in AdS space. The resulting corrections are UV finite, but still visible at distances smaller than the AdS length. We have explored the possible existence of a nontrivial IR structure of the theory associated with the boundary conditions, by determining an $a$-function through the $R$-dependence of the entropy along the lines of \cite{Abate:2026apg}. This analysis revealed the presence of only the UV fixed point for $R\to 0$, associated with the logarithmic term with coefficient $-1/90$ discussed above. 

A natural extension of this work is the study of entangling surfaces anchored on the AdS boundary \cite{Sugishita:2016iel} in order to establish a direct connection with the Ryu-Takayanagi conjecture \cite{Ryu:2006bv,Ryu:2006ef} and provide further insight into the holographic interpretation of entanglement entropy. Our improved understanding of the role of boundary conditions is an important 
step in this direction.

\acknowledgments

We would like to thank
I. Papadimitriou for very useful discussions. 

\appendix
\section{Analytical calculation of the entropy in $1+1$ dimensions}\label{appendix}

In this appendix we calculate analytically the entanglement entropy for the $(1+1)$-dimensional massless scalar theory on a flat background, for boundary conditions that are Dirichlet on the left boundary and Neumann on the right. This problem corresponds to the contribution to the entropy arising from the $\ell=0$ sector of the conformally coupled scalar theory in $3+1$ dimensions, as can be seen by setting $\ell=0$ in~\eqref{eq:Hlm}. We calculate the entanglement entropy of a subsystem $[0,R)$ of a larger system $[0,L]$ in the limit $R\ll L$. The correspondence with AdS$_4$ is achieved for $L/a={\pi}/{2} $. We set $a=1$ in the following.

\subsection{The kernels}

First, we define the kernel $\Omega(w,w^\prime)$, which corresponds to the positive square root of the coupling matrix (\ref{eq:coupling matrix}) in the continuum limit. For the above boundary conditions,
%\subsection{The Kernel $\Omega(w,w^\prime)$}
%The kernel $\Omega(w,w^\prime)$ 
it is defined by the following infinite sum and closed-form expression:
\begin{align}
	\Omega(w,w^\prime) &= \sum_{n=0}^{\infty} \frac{2}{L} \frac{\pi}{L} \left(n + \frac{1}{2}\right) \sin\left[ \left(n + \frac{1}{2}\right) \frac{\pi w}{L} \right] \sin\left[ \left(n + \frac{1}{2}\right) \frac{\pi w^\prime}{L} \right] \nonumber \\
	&= \frac{\pi}{4 L^2} \left( \frac{\cos\frac{\pi (w + w^\prime)}{2 L}}{\sin^2 \frac{\pi (w + w^\prime)}{2 L} } - \frac{\cos\frac{\pi (w - w^\prime)}{2 L}}{\sin^2\frac{\pi (w - w^\prime)}{2 L}} \right).
\end{align}
Its inverse $\Omega^{-1}(w,w^\prime)$ is
\begin{align}
	\Omega^{-1}(w,w^\prime) &= \frac{2}{L} \sum_{n=0}^{\infty} \frac{1}{\left(n + \frac{1}{2}\right) \frac{\pi}{L}} \sin\left[ \left(n + \frac{1}{2}\right) \frac{\pi w}{L} \right] \sin\left[ \left(n + \frac{1}{2}\right) \frac{\pi w^\prime}{L} \right] \nonumber \\
	&= \frac{1}{2 \pi} \ln \frac{\tan^2 \frac{\pi (w + w^\prime)}{4 L}}{\tan^2\frac{\pi (w - w^\prime)}{4 L}}.
\end{align}
%\subsection{Composition of the Kernels}

For $y\neq w$ and $y\neq w^\prime$, the product of the two kernels can be written as a derivative with respect to $y$:
\begin{equation}
	\Omega^{-1}(w,y) \Omega(y,w^\prime) = \frac{2}{L \pi} \frac{\partial}{\partial y} \left( I_1 + I_2 + I_3 \right),
	\label{eq:composition}
\end{equation}
where 
\begin{align}
	I_1 &= -\frac{\cos\frac{\pi w}{2 L} \sin\frac{\pi w^\prime}{2 L}}{\cos\frac{\pi w}{L} - \cos\frac{\pi w^\prime}{L}} \ln \frac{\sin\frac{\pi \vert w - y\vert}{2 L}}{\sin\frac{\pi (w + y)}{2 L}} ,\\
	I_2 &= \frac{\cos\frac{\pi y}{2 L} \sin\frac{\pi w^\prime}{2 L}}{\cos\frac{\pi y}{L}- \cos\frac{\pi w^\prime}{L}} \ln \frac{\tan \frac{\pi \vert w - y \vert}{4 L}} {\tan\frac{\pi (w + y)}{4 L}} ,\\
	I_3 &= \frac{\cos\frac{\pi w^\prime}{2 L} \sin\frac{\pi w}{2 L}}{\cos\frac{\pi w}{L} - \cos\frac{\pi w^\prime}{L}} \ln \frac{\sin\frac{\pi \vert w^\prime - y\vert}{2 L}}{\sin\frac{\pi (w^\prime + y)}{2 L}}.
\end{align}

\subsection{The entanglement entropy}

The calculation of entanglement entropy uses the eigenvalues of the matrix $\mathcal{M}$, defined in equation~\eqref{eq:mmatrix}. In the continuum limit, this matrix becomes the kernel
\begin{equation}
	\mathcal{M}(w,w') = \int_{0}^{R} dy \, \Omega^{-1}(w,y) \Omega(y,w'),
\end{equation}
where we assumed that the subsystem $C$ consists of the degrees of freedom in the interval $[0, R)$\footnote{In the appendix we consider that we trace out the exterior of the sphere, whereas in the main text we trace out the interior. The entanglement entropy is identical in the two cases, since the overall system lies in a pure state.}. Besides $y$, both $w$ and $w'$ take values in this interval. It is advantageous to extend the range of integration to $L$ in order to isolate a $\delta$-function contribution.\footnote{By definition the kernels $\Omega$ and $\Omega^{-1}$ obey $\int_{0}^{L} dy \, \Omega^{-1}(w,y) \Omega(y,w') = \delta(w-w')$.} For this reason, we write
\begin{equation}
	\mathcal{M}(w,w') = \delta(w-w') - \tilde{\mathcal{M}}(w,w'),
\end{equation}
where
\begin{equation}
	\tilde{\mathcal{M}}(w,w') = \int_{R}^{L} dy \, \Omega^{-1}(w,y) \Omega(y,w').
	\label{eq:M_tilde_def}
\end{equation}
Obviously $\mathcal{M}$ and $\tilde{\mathcal{M}}$ share the same eigenfunctions and their eigenvalues are related via $\lambda = 1 - \tilde{\lambda}$. Thus, equation~\eqref{eq:entropylambda} assumes the form
\begin{equation}
	S_{EE} = \sum_{i} \left( \frac{\sqrt{1-\tilde{\lambda}_i}+1}{2} \ln \frac{\sqrt{1-\tilde{\lambda}_i}+1}{2} - \frac{\sqrt{1-\tilde{\lambda}_i}-1}{2} \ln \frac{\sqrt{1-\tilde{\lambda}_i}-1}{2} \right),
	\label{eq:entropyeig}
\end{equation}
where $\tilde{\lambda}_i$ are the eigenvalues of $\tilde{\mathcal{M}}$. Notice that, unless we impose boundary conditions on the eigenfunctions, the spectrum is continuous and the summation must be replaced by an integration.

Substituting equation~\eqref{eq:composition} in~\eqref{eq:M_tilde_def}, we obtain
\begin{multline}
	\tilde{\mathcal{M}}(w,w') = \frac{2}{L \pi}\left(\frac{\cos\frac{\pi w}{2 L} \sin\frac{\pi w^\prime}{2 L}}{\cos\frac{\pi w}{L} - \cos\frac{\pi w^\prime}{L}} \ln \frac{\sin\frac{\pi (R - w)}{2 L}}{\sin\frac{\pi (R + w)}{2 L}}  -\frac{\cos\frac{\pi R}{2 L} \sin\frac{\pi w^\prime}{2 L}}{\cos\frac{\pi R}{L}- \cos\frac{\pi w^\prime}{L}} \ln \frac{\tan \frac{\pi (R -w)}{4 L}} {\tan\frac{\pi (R + w)}{4 L}} \right. \\ \left.-\frac{\cos\frac{\pi w^\prime}{2 L} \sin\frac{\pi w}{2 L}}{\cos\frac{\pi w}{L} - \cos\frac{\pi w^\prime}{L}} \ln \frac{\sin\frac{\pi (R - w^\prime)}{2 L}}{\sin\frac{\pi (R + w^\prime)}{2 L}} \right).
\end{multline}

\subsection{Perturbative expansion in $R/L$}

For $R\ll L={\pi}/{2}$, the kernel $\tilde{\mathcal{M}}(w,w')$ admits an expansion of the form
\begin{equation}
	\tilde{\mathcal{M}}(w,w') = \sum_{i=0}^{\infty} \delta^i \tilde{\mathcal{M}}^{(i)}(w,w'),
\end{equation}
where the expansion parameter is defined as $\delta = {\pi^2 R^2}/{L^2}=4R^2$. The first two terms of this expansion are
\begin{align}
	\tilde{\mathcal{M}}^{(0)} &= \frac{2}{\pi^2} \frac{1}{w^2 - {w^\prime}^2} \left( w^\prime \frac{R^2 - w^2}{R^2 - {w^\prime}^2} \ln \frac{R + w}{R - w} - w \ln \frac{R + w^\prime}{R - w^\prime} \right) ,\\
	\tilde{\mathcal{M}}^{(1)} &= -\frac{1}{12 \pi^2} \frac{w}{R^2} \left( \frac{2 R w^\prime}{R^2 - {w^\prime}^2} + \ln \frac{R + w^\prime}{R - w^\prime} \right).
\end{align}

By the same token, the entanglement entropy, given by equation~\eqref{eq:entropyeig}, is expanded as
\begin{equation}
	S_{EE} = \sum_{i=0}^{\infty} \delta^i S_{EE}^{(i)}.
\end{equation}
It is straightforward to show that
\begin{equation}
	S_{EE}^{(0)} = \sum_{i} \left( \frac{\sqrt{1-\tilde{\lambda}_i^{(0)}}+1}{2} \ln \frac{\sqrt{1-\tilde{\lambda}_i^{(0)}}+1}{2}- \frac{\sqrt{1-\tilde{\lambda}_i^{(0)}}-1}{2} \ln \frac{\sqrt{1-\tilde{\lambda}_i^{(0)}}-1}{2} \right)
\end{equation}
and
\begin{equation}
	S_{EE}^{(1)} = -\sum_{i} \frac{\tilde{\lambda}_i^{(1)}}{2 \sqrt{1-\tilde{\lambda}_i^{(0)}}} \operatorname{arccoth}\sqrt{1-\tilde{\lambda}_i^{(0)}},
	\label{eq:S1}
\end{equation}
where $\tilde{\lambda}_i^{(0)}$ are the eigenvalues of $\tilde{\mathcal{M}}$ at zeroth order in $\delta$, i.e., the eigenvalues of $\tilde{\mathcal{M}}^{(0)}$, and $\tilde{\lambda}_i^{(1)}\delta$ are their leading order corrections.
% Obviously, $S_{EE}^{(0)}$ is the entanglement entropy of a massless theory in flat space, whereas $S_{EE}^{(1)}\,\delta$ is the leading correction due to the curvature of the background.

\subsection{The zeroth-order problem}

The zeroth-order eigenvalue problem is discussed in \cite{Katsinis:2024gef}, see also \cite{Callan:1994py}. Adapting the notation to match that of our calculation, the right eigenfunctions $f(w;\omega)$, the left eigenfunctions $g(w;\omega)$, and the eigenvalues $\tilde{\lambda}^{(0)}(\omega)$ of $\tilde{\mathcal{M}}^{(0)}(w,w')$ read
\begin{align}
	f(w;\omega) &= \sin (\omega u(w)), 
	\\
	g(w;\omega) &= \frac{1}{R} \cosh^2\frac{ u(w) }{2} \sin (\omega u(w)), 
	\\
	\tilde{\lambda}^{(0)}(\omega) &= -\frac{1}{\sinh^2(\pi\omega)},
\end{align}
where
\begin{align}
	u(w) &= \ln \frac{R + w}{R - w}, 
	\label{eq:changevar1}
	\\
	w(u) &= R \tanh\frac{u}{2}.
	\label{eq:changevar2}
\end{align}
The eigenfunctions are normalized according to
\begin{equation}
	\int_{0}^{R} dw \, f(w;\omega) g(w;\omega') = \frac{\pi}{4} \delta(\omega-\omega').
\end{equation}

The leading-order correction to the eigenvalues, $\tilde{\lambda}^{(1)}(\omega)$, are given by
\begin{equation}
	\tilde{\lambda}^{(1)}(\omega) = \frac{\int_{0}^{R} dw \int_{0}^{R} dw' \, \tilde{\mathcal{M}}^{(1)}(w,w') f(w';\omega) g(w;\omega)}{\int_{0}^{R} dw \, f(w;\omega) g(w;\omega)}.
\end{equation}
The denominator is divergent, and the integral must be regularized. This is why this factor cannot be absorbed in a redefinition of the eigenfunctions $f$ and $g$. In particular, using the change of variable~\eqref{eq:changevar2} we obtain
\begin{equation}
	\int_{0}^{R} dw \, f(w;\omega) g(w;\omega) = \frac{1}{2} \int_{0}^{\infty} du \, \sin^2 (\omega u).
\end{equation}
One natural way to regularize the integral is to restrict the integration over $w$ up to $R - \epsilon$, where $\epsilon$ is a UV regulator. Thus, we do not integrate all the way up to the entangling surface, but we stop infinitesimally before it. This regularization scheme is appropriate for dealing with the strong entanglement between adjacent degrees of freedom separated by the entangling surface. With respect to the coordinate $u$, this procedure is equivalent to introducing an upper limit of integration $u_{\max}$. Equation~\eqref{eq:changevar1} implies that the two regulators $\epsilon$ and $u_{\max}$ are related by
\begin{equation}
	u_{\max} = \ln \frac{2 R}{\epsilon}.
\end{equation}
Therefore, we obtain
\begin{equation}
	\int_{0}^{R-\epsilon} dw \, f(w;\omega) g(w;\omega) = \frac{1}{4} u_{\max}.
\end{equation}
Moreover, the UV scale can be used in order to discretize the spectrum according to
\begin{equation}
	\omega_k = \frac{k \pi}{u_{\max}}, \quad k \in \mathbb{N}^*.
	\label{eq:omega_disp}
\end{equation}

\subsection{The correction to the eigenvalues}

The correction to the eigenvalues takes the explicit form
\begin{equation}
	\tilde{\lambda}^{(1)}(\omega) = -\frac{1}{6 \pi^2 u_{\max}} \int_{0}^{\infty}du\int_{0}^{\infty}du^{\prime} \sin (\omega u) \tanh \frac{u}{2} \left( \tanh \frac{u^\prime}{2} + \frac{u^\prime}{2 \cosh^2 \frac{u^\prime}{2}} \right) \sin (\omega u^\prime).
\end{equation}
%\begin{equation}
%	\tilde{\lambda}^{(1)}(\omega) = -\frac{4}{6 \pi^2 u_{\max}} %\int_{0}^{\infty}du\int_{0}^{\infty}du^{\prime} \sin (2\omega u) \tanh %u\left( \tanh u^\prime+ \frac{u^\prime}{\cosh^2 u^\prime} \right) \sin %(2\omega u^\prime)
%\end{equation}
To compute the integrals, we employ the master contour integral \cite{gradshteyn2007table}
\begin{equation}
	\int_{-\infty}^{\infty} d\omega \, \frac{\sinh (a\omega)}{\cosh (b\omega)} e^{i\omega u^\prime} = i \frac{2\pi}{b} \frac{\sin\frac{\pi a}{2b} \sinh\frac{\pi u^\prime}{2b}}{\cos\frac{\pi a}{b} + \cosh\frac{\pi u^\prime}{b}}.
\end{equation}
Making use of the imaginary component of the Fourier transform  for $a=b=1$, we obtain
\begin{equation}
	\int_{0}^{\infty} du \, \tanh u \sin (\omega u) = \frac{\pi}{2 \sinh \frac{\pi\omega}{2}}.
\end{equation}
The integral involving the derivative of the hyperbolic tangent (the squared hyperbolic secant component) evaluates to 
\begin{equation}
	\int_{0}^{\infty} du \, \frac{u \sin \omega u}{\cosh^2 u} = \frac{\pi}{4} \left( -2 + \pi \omega \coth \frac{\pi \omega}{2} \right) \csch \frac{\pi \omega}{2}=\frac{\pi}{4} \frac{\pi \omega \coth \frac{\pi \omega}{2} - 2}{\sinh \frac{\pi \omega}{2}}.
\end{equation}
%
%Alternatively, writing the hyperbolic cosecant explicitly in terms of %the hyperbolic sine yields:
%\begin{equation}
%	\int_{0}^{\infty} du \, \frac{u \sin \omega u}{\cosh^2 u} = %\frac{\pi}{4} \frac{\pi \omega \coth \frac{\pi \omega}{2} - 2}{\sinh %\frac{\pi \omega}{2}}
%\end{equation}
%
Using these integrals we obtain a remarkably compact expression, which
can be rewritten as a parametric derivative of the zeroth-order result:
\begin{equation}
	\tilde{\lambda}^{(1)}(\omega) = -\frac{1}{6u_{\max}} \frac{\pi \omega \cosh (\pi \omega)}{\sinh^3 (\pi \omega)}=-\frac{\omega}{12 u_{\max}} \frac{d}{d\omega} \tilde{\lambda}^{(0)}(\omega) .
\end{equation}

\subsection{The correction to the entropy}

Substituting in equation~\eqref{eq:S1} we obtain
\begin{equation}
	S_{\mathrm{EE}}^{(1)} = \frac{1}{12u_{\max}}\sum_{i} \frac{\pi^2 \omega_i^2}{\sinh^2 (\pi \omega_i)},
\end{equation}
where the frequencies are given by equation~\eqref{eq:omega_disp}. For $u_{\max}\rightarrow\infty$ we approximate the sum with the integral
\begin{equation}
	S_{\mathrm{EE}}^{(1)} = \frac{1}{12\pi^2}\int_{0}^{\infty}d\omega \frac{\omega^2}{\sinh^2 \omega}=\frac{1}{72}.
\end{equation}
Thus, for Dirichlet-Neumann boundary conditions at the two ends of the total system, the entanglement entropy is
\begin{equation}
	S_{\mathrm{EE}}\equiv S_{\mathrm{DN}} = S_{\mathrm{EE}}^{(0)}+S_{\mathrm{EE}}^{(1)}\,\delta+\dots=
	\frac{1}{6}\ln\left(\frac{2R}{\epsilon}\right)+
	\frac{1}{18} R^2 + \dots
	\label{eq:entropy1ddn}
\end{equation}
On the other hand, for Dirichlet-Dirichlet boundary conditions, the well-known result gives \cite{Calabrese:2004eu}
\begin{equation} 
	S_{\mathrm{DD}}=\frac{1}{6}\ln\left(\frac{2L}{\pi \epsilon}\sin\frac{\pi R}{L}\right)=	\frac{1}{6}\ln\left(\frac{2R}{\epsilon}\right)-\frac{1}{9} R^2 
	+ \dots
	\label{eq:entropy1ddd}
\end{equation}
for $L={\pi}/{2}$.
The difference in the boundary conditions at the right end of the total system results in
\begin{equation}
	\begin{split}
		\Delta S = S_{\mathrm{DN}}-S_{\mathrm{DD}}= \frac{1}{6} R^2  + \dots
	\end{split}
\end{equation}

%\bibliographystyle{JHEP} % or JHEP.bst if available
%\bibliography{AdS_bib}

\end{document}